\documentclass[oneside,12pt]{article}
\usepackage{amsmath}
\usepackage{amsfonts}
\usepackage{graphicx,epsfig}
\usepackage{eps to pdf}
\usepackage{cite}
\usepackage{amssymb}

\begin{document}

\title{\textbf{\LARGE Reconstruction of cosmic history from a simple
parametrization of $H$}}
\author{{\large S. K. J. Pacif}$^{1}${\large , R. Myrzakulov}$^{2}$, {\large %
S. Myrzakul}$^{3}$ \\
%EndAName
$^{1}$\textit{Centre for Theoretical Physics,}\\
\textit{\ \ Jamia Millia Islamia,}\\
\textit{\ \ New Delhi 110025, India}\\
\textit{\ }$^{2}$\textit{Eurasian International Center for Theoretical
Physics,}\\
\textit{\ Eurasian National University, Astana 010008, Kazakhstan}\\
$^{3}$\textit{Department of Theoretical and Nuclear Physics,}\\
\textit{\ Al-Farabi Kazakh National University, Al-Farabi av. 71, 050040, \ }%
\\
\textit{Almaty, Kazakhstan}\\
\textit{\ \ shibesh.math@gmail.com}$^{1}$\textit{, rmyrzakulov@gmail.com}$%
^{2}$, \ \\
\ \ \textit{shynaray1981@gmail.com}$^{3}$\\
}
\maketitle

\begin{abstract}
In this paper, we propose a simple parametrization of the Hubble parameter $%
H $ in order to explain the late time cosmic acceleration. We show that our
proposal covers many models obtained in different schemes of parametrization
under one umbrella. We demonstrate that a simple modification in the
functional form of Hubble parameter can give rise to interesting
cosmological phenomena such as big rip singularity, bounce and others. We
have also constrained the model parameters using the latest 28 points of $%
H(z)$ data for three cases which admit transition from deceleration to
acceleration.
\end{abstract}


\section{\protect\Large Introduction}

\qquad One of the aims of cosmology is to determine a mathematical model of
the large scale structure of the Universe which can explain the results of
astronomical observations and whose dynamics can be determined by the
physical laws describing the behavior of matter on larger scale. According
to Alan Sandage \cite{sandage}, in cosmology at the background level, one
search for two numbers: $H_{0}$ and $q_{0}$ (suffix '$0$' stands for the
present value of the quantity), where $H$ (Hubble parameter (HP)) and $q$
(deceleration parameter(DP)) are two dynamical quantities which tell about
the expansion rate of the Universe. But the present day cosmology use around
four to twenty parameters to explain the Universe. Still, $H$ and $q$ play
the central role in the Einstein's field equations (EFEs) explaining the
observations. They can be defined naturally in the linear and non linear
derivatives of scale factor $a(t)$ in the Taylor series expansion of $a(t)$
in the vicinity of the present time $t_{0}$ as%
\begin{equation}
a(t)=a\left( {{t}_{0}}\right) +\dot{a}\left( {{t}_{0}}\right) \left[ t-{{t}%
_{0}}\right] +\frac{1}{2}\ddot{a}\left( {{t}_{0}}\right) {{\left[ t-{{t}_{0}}%
\right] }^{2}}+\cdots \text{.}  \label{SF1}
\end{equation}%
An overhead dot `$\cdot $' represents derivative w.r.t. cosmic time `$t$'.
From equation (\ref{SF1}), we obtain 
\begin{equation}
\frac{a(t)}{a\left( t_{0}\right) }=1+H_{0}\left[ t-t_{0}\right] -\frac{q_{0}%
}{2}H_{0}^{2}{{\left[ t-{{t}_{0}}\right] }^{2}}+\cdots ,  \label{SFT2}
\end{equation}%
where\qquad

\begin{equation}
H(t)=\frac{\dot{a}}{a}\text{ , }q(t)=-\frac{a\ddot{a}}{\dot{a}^{2}}\text{.}
\label{H q}
\end{equation}%
\qquad \qquad

Till today the most successful theory explaining the Universe is the big
bang theory which is based on general relativity. After Hubble's work,
cosmologists made attempts to measure the deceleration of the expansion with
the belief that the expansion of the Universe must slow down caused by
gravity. However, the observations of distant supernovae of type Ia by
Supernova Cosmology Project \cite{SCP} and the High-Z Supernova Search team 
\cite{HZTEAM} gave totally unexpected result to the fact that the expansion
of the Universe is accelerating. Since then further searches presented
convincing evidence for accelerating expansion with greater accuracy \cite%
{HZ1}, \cite{HZ2}, \cite{HZ3}, \cite{SCP1}, \cite{SCP2}, \cite{SCP3}. The
fact is also supported by some other observations such as CMB \cite{cmb1}, 
\cite{cmb2}, BAO \cite{bao1}, \cite{bao2}, SDSS \cite{sdss1}, \cite{sdss2}
etc. For both cosmology and physics, the cosmic acceleration is probably an
important discovery. It raised a lot of questions on the fundamental
principles funded with cosmology. Based on the accelerating expansion of the
Universe, the past few years produced a plethora of cosmological models
either by modifying the energy momentum tensor in the right hand side of
Einstein's field equation (EFE) or by modifying the gravity theory
(modifying the LHS of EFE). Alternative theories are also there such as the
inclusion of inhomogeneity, back reaction, averaging etc. Recently, a series
of papers by Vishwakarma \cite{vishwa0}, \cite{vishwa1}, \cite{vishwa2}, 
\cite{vishwa3} explained this fact in a simple and viable way which raised
some questions on the geometrization of gravity theory.

Although there are several ways to describe this cosmic acceleration, but it
is generally attributed to the presence of dark energy (DE) throughout the
Universe. Obviously it gives rise to the question of what this mysterious DE
really is, what is its nature and why it starts dominating the Universe so
recently.\ The literature contains numerous models of DE but the simplest
and popular candidate of DE is the Einstein's cosmological constant ($%
\Lambda $) \cite{sahniCC}, \cite{vishwaCC}, \cite{peeblesCC}. However, it
suffers from the well known cosmological constant problem \cite{WEINBERG}
which can be alleviated by considering a dynamically decaying $\Lambda $. On
the other hand primordial inflation has taken a special status in explaining
the origin of the anisotropies in the CMB radiation and the formation of
large scale structures. This motivates theorists to invoke scalar field to
explain the early and the late time acceleration together. So far, a wide
variety of scalar field models of DE have been proposed in the past few
years including quintessence \cite{quint1}, \cite{quint2}, \cite{quint3}, 
\cite{quint4}, K-essence \cite{kessen1}, \cite{kessen2}, spintessence \cite%
{spint}, tachyon \cite{tachy1}, \cite{tachy2}, quintom \cite{quintom1}, \cite%
{quintom2}, \cite{quintom3}, \cite{quintom4}, chameleon \cite{chamel1}, \cite%
{chamel2}, \cite{chamel3} and many more. Though these scalar field models
give the equation of state (EoS) parameter $\left( w=\frac{p}{\rho }\right) $%
, $-\frac{1}{3}<w<0$, observational data also allow models of DE with EoS
parameter crossing $-1$ line (called phantom field models). A number of
phantom field models have been proposed \cite{phant1}, \cite{phant2}, \cite%
{phant3}, \cite{phantsami}, \cite{phant4}, \cite{phant5} in the past few
years. Another way to explain the acceleration is to incorporate the
Chaplygin gas \cite{CG1}, \cite{CG2} into the EFEs. For a brief review on
candidates of DE and alternatives to DE, one can see \cite{DEREV1}, \cite%
{DEREVa}, \cite{DEREVb}, \cite{DEREVc}, \cite{DEREVd}, \cite{DEREVe}, \cite%
{DEREV2}.

Observations suggest that the cosmic acceleration is a recent phenomena and
the Universe has entered a state of accelerating expansion around redshift $%
z\sim 0.5$. The existence of decelerated expansion phase in the Universe is
also supported by the gravitational instability theory of structure
formation and of big bang nucleosynthesis. This implies that the Universe
must have undergone from decelerated to accelerated phase of evolution. This
motivates the theorists in modeling the Universe with deceleration to
acceleration phase transition. The kinematic approach is discussed in \cite%
{Turner} to explain the cosmic acceleration which do not assume the validity
of general relativity or any particular gravitational theory (see \cite%
{Weinberg}). This method do not effect the physical or geometrical
properties of DE and is known as the model-independent way to study the DE
i.e. by a parametrized EoS of dark energy. For a review on parametrization
of equation of state parameter $w$, one can see \cite{LINDER-w}. Another
model-independent way to study the DE is by parametrizing the DP. For a
brief review on DP, one can see \cite{BOLOTIN-q}, \cite{BOLO}. Also, there
are several parametrization of HP considered by Nojiri and Odintsov (and
collaborators) \cite{H-odin3} to study the future cosmological
singularities. Here, in this paper, we have studied the evolution of the
Universe by parametrizing the functional form of $H$ and see how it reduces
to some known obtained models and explains the late time cosmic acceleration.

\section{Motivation}

\qquad We would like to stress on the cosmographic parameter $H$ describing
the expansion of the Universe and its role in generating some interesting
mathematical models of the Universe in Einstein's theory of gravitation. In
FRW cosmology, there are three variables namely $a(t),$ $\rho (t),$ $p(t)$
with two independent equations which can be solved by supplementing the
equation of state(s), $w=\frac{p}{\rho }$ of the energy component(s). In
this case the parameter $w$ is a constant. For a dynamical $\Lambda $, one
needs one more constrain equation to close the system. This extra constrain
equation (or the supplementary equation) has been chosen in various ways in
order to explain the standard cosmological problems such as to solve the
cosmological constant problem, singularity problem, horizon problem,
flatness problem, density fluctuation problem, dark matter problem, exotic
relics problem, thermal state problem etc. during the past forty-fifty
years. With the addition of the DE component into the field equations the
EoS parameter $w$ becomes dynamical ($w(t)$). Thus, there are many
traditional ways to choose this supplementary equation relating any two
variables involved in the field equations e.g. $p\sim \rho $, $\Lambda \sim
a^{-2},\ \Lambda \sim \rho $ etc. Also, one can parametrize any variable to
get this extra constrain equation to close the system. The various
parametrization used in literature relating to $a(t),$ $\rho (t),$ $p(t),$ $%
\Lambda (t),$ $q(t)$ or $w(t)$ are summarized here in detail (see table-6 to
table-11 in appendix-1).

Also there are various parametrization of Hubble parameter $H(t)$ in
literature used to explain some problems of standard cosmology and are
listed in the below table-1.

\begin{tabular}{lc}
{\small Table-1} &  \\ \hline
\multicolumn{1}{|l}{{\small Parametrization of HP (}$H${\small )}} & 
\multicolumn{1}{|c|}{\small Ref.} \\ \hline
\multicolumn{1}{|l}{$H(a)=Da^{-m}${\small \ (}$D${\small \ and }$m${\small \
are constants)}} & \multicolumn{1}{|c|}{{\small \cite{H-BermanNuovo}}} \\ 
\hline
\multicolumn{1}{|l}{$H(a)=e^{\frac{1-\gamma a^{2}}{\alpha a}}${\small \ (}$%
\gamma ${\small \ and }$\alpha ${\small \ are constants)}} & 
\multicolumn{1}{|c|}{{\small \cite{H-Banerjee1}}} \\ \hline
\multicolumn{1}{|l}{$H(a)=\alpha (1+a^{-n})${\small \ (}$\alpha ${\small \
and }$n${\small \ are constants)}} & \multicolumn{1}{|c|}{{\small \cite%
{H-JPSINGH}}} \\ \hline
\multicolumn{1}{|l}{$H(t)=\frac{m}{\alpha t+\beta }${\small \ (}$\alpha ,$%
{\small \ }$\beta ,${\small \ }$m${\small \ are constants)}} & 
\multicolumn{1}{|c|}{{\small \cite{H-Pacif1}, \cite{H-Pacif2}}} \\ \hline
\multicolumn{1}{|l}{$H(t)=\frac{\alpha t_{R}}{t(t_{R}-t)}${\small \ (}$%
\alpha ${\small \ is a constant, }$t_{R}${\small \ is big Rip time)}} & 
\multicolumn{1}{|c|}{{\small \cite{H-Cannata}}} \\ \hline
\multicolumn{1}{|l}{${\small H(t)=}\frac{\alpha }{3}\left( t+T_{0}\right)
^{3}{\small -\beta }\left( t+T_{0}\right) {\small +\gamma ,}$} & 
\multicolumn{1}{|c|}{{\small \cite{H-nojiri}, \cite{chamel3}}} \\ 
\multicolumn{1}{|l}{$\gamma =-\frac{\alpha }{3}T_{0}^{3}+\beta T_{0}${\small %
\ (}$\alpha ,${\small \ }$\beta ,${\small \ }$T_{0}${\small \ are constants)}
} & \multicolumn{1}{|c|}{} \\ \hline
\multicolumn{1}{|l}{$H(t)=H_{0}e^{\lambda t}${\small \ (}$H_{0},${\small \ }$%
\lambda ${\small \ are constants)}} & \multicolumn{1}{|c|}{{\small \cite%
{H-odinF}}} \\ \hline
\multicolumn{1}{|l}{$H(t)=c_{0}+b_{0}(t_{s}-t)^{\alpha }${\small \ (}$%
c_{0},b_{0},\alpha ${\small \ are constants)}} & \multicolumn{1}{|c|}{%
{\small \cite{H-odin1}}} \\ \hline
\multicolumn{1}{|l}{$H(t)=H_{0}-H_{1}e^{-\beta t}${\small \ (}$H_{0}>0,$%
{\small \ }$H_{1}>0,${\small \ }$\beta ${\small \ are constants)}} & 
\multicolumn{1}{|c|}{{\small \cite{H-odin2}}} \\ \hline
\multicolumn{1}{|l}{${\small H(t)=f}_{1}{\small (t)+f}_{2}{\small (t)(t}_{s}%
{\small -t)}^{\alpha }$} & \multicolumn{1}{|c|}{{\small \cite{H-odin3}}} \\ 
\multicolumn{1}{|l}{{\small (}$f_{1}(t)${\small \ \& }$f_{2}(t)${\small \
are arbitrary functions, }$\alpha ${\small \ constant)}} & 
\multicolumn{1}{|c|}{} \\ \hline
\end{tabular}

One can find some more parametrization of HP in \cite{H-odin3}.

From equation (\ref{H q}), we find

\begin{equation}
a(t)=Ce^{\int H(t)dt}\text{, where }C\text{ is a constant of integration.}
\label{A1}
\end{equation}

\begin{equation}
q(t)=-1+\frac{d}{dt}\left( \frac{1}{H(t)}\right) ,  \label{A2}
\end{equation}

EFEs can also be expressed as

\begin{equation}
\sum \rho _{i}(t)=3M_{Pl}^{2}\left[ \left( H(t)\right) ^{2}+\frac{k}{a^{2}}%
\right] ,  \label{A3}
\end{equation}

\begin{equation}
\sum \left[ 1+3w_{i}(t)\right] \rho _{i}(t)=6M_{Pl}^{2}\left( H(t)\right)
^{2}\left[ -1+\frac{d}{dt}\left( \frac{1}{H(t)}\right) \right] .  \label{A4}
\end{equation}%
\qquad

In the above equations all the physical variables are in terms of $H(t)$.
Now, it is easy to see that a simple integrable form of $H(t)$ will
determine all the physical variables smoothly. We prefer to parametrize the
HP because the variation of Hubble's law assumed is not inconsistent with
observations and has the advantage of providing simple functional form of
the time evolution of the scale factor and so as dynamics. Motivated by the
above discussions, we propose a simple and convenient form of HP as an
explicit function of cosmic time `$t$' in the form%
\begin{equation}
H(t)=\frac{\beta t^{m}}{\left( t^{n}+\alpha \right) ^{p}}  \label{mainansatz}
\end{equation}%
where $\alpha ,$ $\beta \neq 0,$ $m,$ $n,$ $p$ are real constants (better
call them model parameters). $\alpha $ and $\beta $ both have the dimensions
of time. The specific values of $m,$ $n,$ $p$ will suggest the different
forms of HP and produce interesting cosmologies. Our parametrization
generalizes several known models which were obtained by the parametrization
of any cosmological parameters $a(t),$ $H(t),$ $q(t),$ $\Lambda (t),$ $\rho
(t)$ or $w(t)$ in different contexts. In the next section, we formulate the
Einstein's field equations for a general scalar field cosmology and solve
the system with the help of our main ansatz.

\section{Field equations and solutions}

\qquad We know scalar fields are extremely important in modern physics being
invariant under coordinate transformations. There have been a great activity
in modelling the Universe with a motivation to explain both the early and
late time acceleration of the Universe with scalar fields. We know, the
nature of DE remain matters of speculation, but it is generally believed to
be homogeneous, not very dense and is not known to interact through any of
the fundamental forces other than gravity. So, it can be represented as
large scale scalar field $\phi $. For an ordinary scalar field $\phi $\
minimally coupled to gravity with Lagrangian density $\mathcal{L}=-\frac{1}{2%
}g^{\mu \nu }\partial _{\mu }\phi \partial _{\nu }\phi -V(\phi )$, the
action is given by

\begin{equation}
S=\int d^{4}x\sqrt{-g}\left[ -\frac{1}{2}g^{\mu \nu }\partial _{\mu }\phi
\partial _{\nu }\phi -V\left( \phi \right) \right] ,  \label{ACT}
\end{equation}%
where $V\left( \phi \right) $ is the potential of the field. The
stress-energy tensor of the field $\phi $ take the form of a perfect fluid
as \cite{vishwaCC} 
\begin{equation}
T_{\mu \nu }^{\phi }=(\rho _{\phi }+p_{\phi })U_{\mu }U_{\nu }+p_{\phi
}~g_{\mu \nu },  \label{TIJDARKENERGY}
\end{equation}%
where the density and pressure of scalar field are expressed as $\rho _{\phi
}=\frac{\dot{\phi}^{2}}{2}+V\left( \phi \right) $ and $p_{\phi }=\frac{\dot{%
\phi}^{2}}{2}-V\left( \phi \right) $, with the understanding that $\phi $ is
spatially homogeneous. The evolution of the scalar field is governed by the
wave equation $\ddot{\phi}+3H\dot{\phi}+V^{^{\prime }}(\phi )=0,$ where a
prime denotes differentiation with respect to $\phi $. The state equation of
scalar field $w_{\phi }$ can be represented as $w_{\phi }=\frac{p_{\phi }}{%
\rho _{\phi }}=\frac{-1+\frac{\dot{\phi}^{2}}{2V}}{1+\frac{\dot{\phi}^{2}}{2V%
}}$. This give rise to several candidates for DE, which depends upon the
dynamics of the field $\phi $ and its potential energy $V\left( \phi \right) 
$. For a slow roll scalar field $\frac{\dot{\phi}^{2}}{2}\ll V\left( \phi
\right) $, it reduces to the case of most favoured cosmological constant $%
\Lambda $ for which $w_{\phi }=-1$. For $-\frac{1}{3}<w_{\phi }<0$ we have
quintessence and $w_{\phi }$ crossing $-1$, phantom field is observed. To
introduce DE into EFEs, we replace the energy momentum tensor $T_{\mu \nu }$
by%
\begin{equation}
T_{\mu \nu }^{Total}=T_{\mu \nu }+T_{\mu \nu }^{\phi }=(\rho
_{Total}+p_{Total})U_{\mu }U_{\nu }+p_{Total}~g_{\mu \nu }  \label{TOTALTIJ}
\end{equation}%
with the understanding that $\rho _{Total}=\rho _{eff}=\rho +\rho _{\phi }$
and $p_{Total}=p_{eff}=p+p_{\phi }$.

With the fluid described here by (\ref{TOTALTIJ}), the EFEs reduce to

\begin{equation}
\rho _{eff}=\rho +\rho _{\phi }=3M_{Pl}^{2}\left( H^{2}+\frac{k}{a^{2}}%
\right) ,  \label{M1}
\end{equation}%
\begin{equation}
p_{eff}=p+p_{\phi }=-M_{Pl}^{2}\left( 2\frac{\ddot{a}}{a}+H^{2}+\frac{k}{%
a^{2}}\right) ,  \label{M2}
\end{equation}%
with the state equations

\begin{equation}
p=w\rho \text{ \ \ \ \ }(0\leqslant w\leqslant 1)\text{ \ \ and \ \ }p_{\phi
}=w_{\phi }\rho _{\phi }.  \label{M3}
\end{equation}

Meanwhile we consider the minimal interaction between matter and dark energy
which yield

\begin{equation}
\dot{\rho}+3H(1+w)\rho =0,\text{ \ }\dot{\rho}_{\phi }+3H(1+w_{\phi })\rho
_{\phi }=0,  \label{M4}
\end{equation}%
leading to $\rho \sim a^{-3(1+w)}$ and $\rho _{\phi }\sim a^{-3(1+w_{\phi
})} $ (for constant $w_{\phi }$ (such as cosmological constant)). But, $%
w_{\phi } $ must be a function of time in general.

With our main ansatz (\ref{mainansatz}) the general expressions for the time
variation of all the CP are obtained as follows

\begin{equation}
a(t)=Ce^{\beta \int \frac{t^{m}dt}{\left( t^{n}+\alpha \right) ^{p}}}\text{, 
}C\text{ is an arbitrary constant of integration,}  \label{Ga}
\end{equation}

\begin{equation}
q(t)=-1+\frac{1}{\beta }\left\{ \left( np-m\right) t^{n}-m\alpha \right\} 
\frac{\left( t^{n}+\alpha \right) ^{p-1}}{t^{m+1}},  \label{Gq}
\end{equation}

\begin{equation}
\rho _{eff}(t)=3M_{Pl}^{2}\left[ \frac{\beta ^{2}t^{2m}}{\left( t^{n}+\alpha
\right) ^{2p}}+\frac{k}{C^{2}}e^{-2\beta \int \frac{t^{m}dt}{\left(
t^{n}+\alpha \right) ^{p}}}\right] ,  \label{Grhototal}
\end{equation}

\begin{equation}
p_{eff}(t)=M_{Pl}^{2}\left[ \left( -3+\frac{2}{\beta }\left\{ \left(
np-m\right) t^{n}-m\alpha \right\} \frac{\left( t^{n}+\alpha \right) ^{p-1}}{%
t^{m+1}}\right) \frac{\beta ^{2}t^{2m}}{\left( t^{n}+\alpha \right) ^{2p}}-%
\frac{k}{C^{2}}e^{-2\beta \int \frac{t^{m}dt}{\left( t^{n}+\alpha \right)
^{p}}}\right] ,  \label{Gptotal}
\end{equation}

\begin{equation}
\rho (t)=\left[ DC^{-3(1+w)}\right] e^{-3(1+w)\beta \int \frac{t^{m}dt}{%
\left( t^{n}+\alpha \right) ^{p}}}\text{, }D\text{ is an arbitrary constant
of integration.}  \label{Grho}
\end{equation}

We observe that some particular values of $m,$ $n,$ $p$ will give explicit
solutions of EFEs. In general there are one, two or three model parameters
in all the parametrization considered (see table-1 and table-6 to table-11).
But, we have five parameters $\alpha ,$ $\beta ,$ $m,$ $n,$ $p$ in the
functional form of HP. Without the loss of generality, we reduce the number
of model parameters by giving some specific values to $m,$ $n,$ $p$ which
will be helpful to analyze the physical and geometrical behavior of our
obtained models. For some suitable choice of integral values of $m,$ $n,$ $p$
(and one non-integral value of $p$), we obtain some specific models leaving $%
\alpha $ and $\beta $ generic. The various models thus obtained with two
model parameters $\alpha $ and $\beta $ are given in the following table-8.
The two model parameters $\alpha $ and $\beta $ can be constrained from any
observational data (e.g. Sne Ia data, H(z) data or BAO data). However, one
can also constrain all five model parameters simultaneously but in this work
we confined to two model parameters $\alpha ,\beta $ by specifying $m,$ $n,$ 
$p$ to see how our parametrization of HP can reproduces some particular
models.

\begin{tabular}{cllll}
Table-2 &  &  &  &  \\ \hline
\multicolumn{1}{|c}{Models} & \multicolumn{1}{|c}{Specific Values of} & 
\multicolumn{1}{|c}{HP} & \multicolumn{1}{|c}{SF} & \multicolumn{1}{|c|}{DP}
\\ 
\multicolumn{1}{|c}{} & \multicolumn{1}{|c}{$m,$\ $n,$\ $p$} & 
\multicolumn{1}{|c}{$H(t)$} & \multicolumn{1}{|c}{$a(t)$} & 
\multicolumn{1}{|c|}{$q(t)$} \\ \hline
\multicolumn{1}{|c}{I} & \multicolumn{1}{|l}{$m=0,p=0,\forall n$} & 
\multicolumn{1}{|l}{$\beta $} & \multicolumn{1}{|l}{$Ce^{\beta t}$} & 
\multicolumn{1}{|l|}{$-1$} \\ \hline
\multicolumn{1}{|c}{II} & \multicolumn{1}{|l}{$m=-1,p=0,\forall n$} & 
\multicolumn{1}{|l}{$\frac{\beta }{t}$} & \multicolumn{1}{|l}{$Ct^{\beta }$}
& \multicolumn{1}{|l|}{$-1+\frac{1}{\beta }$} \\ \hline
\multicolumn{1}{|c}{III} & \multicolumn{1}{|l}{$m=0,p=1,n=1$} & 
\multicolumn{1}{|l}{$\frac{\beta }{t+\alpha }$} & \multicolumn{1}{|l}{$%
C\left( t+\alpha \right) ^{\beta }$} & \multicolumn{1}{|l|}{$-1+\frac{1}{%
\beta }$} \\ \hline
\multicolumn{1}{|c}{IV} & \multicolumn{1}{|l}{$m=1,p=0,\forall n$} & 
\multicolumn{1}{|l}{$\beta t$} & \multicolumn{1}{|l}{$Ce^{\beta \frac{t^{2}}{%
2}}$} & \multicolumn{1}{|l|}{$-1-\frac{1}{\beta }\frac{1}{t^{2}}$} \\ \hline
\multicolumn{1}{|c}{V} & \multicolumn{1}{|l}{$m=0,p=1,n=2$} & 
\multicolumn{1}{|l}{$\frac{\beta }{t^{2}+\alpha }$} & \multicolumn{1}{|l}{$%
Ce^{\frac{\beta }{\sqrt{\alpha }}\tan ^{-1}\frac{t}{\sqrt{\alpha }}}$} & 
\multicolumn{1}{|l|}{$-1+\frac{2}{\beta }t$} \\ \hline
\multicolumn{1}{|c}{VI} & \multicolumn{1}{|l}{$m=0,p=\frac{1}{2},n=1$} & 
\multicolumn{1}{|l}{$\frac{\beta }{\sqrt{t+\alpha }}$} & \multicolumn{1}{|l}{%
$Ce^{2\beta \sqrt{t+\alpha }}$} & \multicolumn{1}{|l|}{$-1+\frac{1}{2\beta }%
\frac{1}{\sqrt{t+\alpha }}$} \\ \hline
\multicolumn{1}{|c}{VII} & \multicolumn{1}{|l}{$m=0,p=\frac{1}{2},n=2$} & 
\multicolumn{1}{|l}{$\frac{\beta }{\sqrt{t^{2}+\alpha }}$} & 
\multicolumn{1}{|l}{$C\left( t+\sqrt{t^{2}+\alpha }\right) ^{\beta }$} & 
\multicolumn{1}{|l|}{$-1+\frac{1}{\beta }\frac{t}{\sqrt{t^{2}+\alpha }}$} \\ 
\hline
\multicolumn{1}{|c}{VIII} & \multicolumn{1}{|l}{$m=1,p=1,n=1$} & 
\multicolumn{1}{|l}{$\frac{\beta t}{t+\alpha }$} & \multicolumn{1}{|l}{$%
Ce^{\beta t}\left( t+\alpha \right) ^{-\alpha \beta }$} & 
\multicolumn{1}{|l|}{$-1-\frac{\alpha }{\beta }\frac{1}{t^{2}}$} \\ \hline
\multicolumn{1}{|c}{IX} & \multicolumn{1}{|l}{$m=1,p=1,n=2$} & 
\multicolumn{1}{|l}{$\frac{\beta t}{t^{2}+\alpha }$} & \multicolumn{1}{|l}{$%
C\left( t^{2}+\alpha \right) ^{\frac{\beta }{2}}$} & \multicolumn{1}{|l|}{$%
-1+\frac{1}{\beta }-\frac{\alpha }{\beta }\frac{1}{t^{2}}$} \\ \hline
\multicolumn{1}{|c}{X} & \multicolumn{1}{|l}{$m=1,p=\frac{1}{2},n=2$} & 
\multicolumn{1}{|l}{$\frac{\beta t}{\sqrt{t^{2}+\alpha }}$} & 
\multicolumn{1}{|l}{$Ce^{\beta \sqrt{t^{2}+\alpha }}$} & 
\multicolumn{1}{|l|}{$-1-\frac{\alpha }{\beta }\frac{1}{t^{2}\sqrt{%
t^{2}+\alpha }}$} \\ \hline
\multicolumn{1}{|c}{XI} & \multicolumn{1}{|l}{$m=-1,p=1,n=1$} & 
\multicolumn{1}{|l}{$\frac{\beta }{t\left( t+\alpha \right) }$} & 
\multicolumn{1}{|l}{$C\left( \frac{t}{t+\alpha }\right) ^{\frac{\beta }{%
\alpha }}$} & \multicolumn{1}{|l|}{$-1+\frac{\alpha }{\beta }+\frac{2}{\beta 
}t$} \\ \hline
\multicolumn{1}{|c}{XII} & \multicolumn{1}{|l}{$m=-1,p=1,n=2$} & 
\multicolumn{1}{|l}{$\frac{\beta }{t\left( t^{2}+\alpha \right) }$} & 
\multicolumn{1}{|l}{$C\left( \frac{t^{2}}{t^{2}+\alpha }\right) ^{\frac{%
\beta }{2\alpha }}$} & \multicolumn{1}{|l|}{$-1+\frac{\alpha }{\beta }+\frac{%
3}{\beta }t^{2}$} \\ \hline
\end{tabular}

\bigskip

In table-2, we see $\Lambda CDM$ model (model-I, where $\beta $ plays the
role of $\Lambda $), power law cosmology (PLC) \cite{PLC1} (model-II),
Berman's model of constant deceleration parameter (BM) \cite{BERMAN1}
(model-III with $\frac{1}{\beta }=m$), Abdel Rahman's model (AR) (model-IX) 
\cite{ABDELMODEL} (with $\beta =1$) and its generalized model \cite{q-ASSRP}
are obtained here. Model-XI imitate the linearly varying deceleration
parameter model (LVDPt) of Akarsu \cite{AKARSU} (where $\frac{2}{\beta }=-k$
and $\frac{\alpha }{\beta }=m$). Thus, we can see all these models come
under our scheme of parametrization of HP for some specific choice of model
parameters. We note that, many solutions obtained here are non-singular
bouncing solutions, where the bounce occur at some finite value of the scale
factor $a(t)$.

For $\alpha =0$, the form of $H(t)$ becomes $H(t)=\beta t^{\gamma }$, ($%
\gamma =(m-np)$ is a new constant) giving the same result as $p=0$
(Models-I,II,IV). The case for negative $\alpha $ is that for which our main
ansatz (\ref{mainansatz}) will take the form $H(t)=\frac{\beta t^{m}}{\left(
t^{n}-\alpha \right) ^{p}}$, $\alpha >0$. In this case the behavior of scale
factor will differ greatly in some models and so as dynamics e.g. with this
form of HP, all the models ($p\neq 0$) have collapsing nature at $t=\alpha ^{%
\frac{1}{n}}$. If we take $H(t)=\frac{\beta t^{m}}{\left( \alpha
-t^{n}\right) ^{p}}$, $\alpha >t$ then this form of HP will lead to models
with future singularity at $t=\alpha ^{\frac{1}{n}}$. One can study these
future singularities for different models so obtained. Modifying the form of
HP to $H(t)=\frac{\beta t^{m}+\eta }{\left( \alpha -t^{n}\right) ^{p}},$ $%
\alpha >t$ and $\eta $ is another parameter then for $p=0$, one can obtain
the hybrid scale factor cosmology \cite{HSF1}, \cite{HSF2} and for $p=-1$
one can obtain some results discussed by Nojiri and Odintsov \cite{H-odin3}.
We should mention here that different evolution of the scale factor with a
variable cosmological term $\Lambda $ is studied extensively by Overduin and
Cooperstock in \cite{COOPER}. Similarly, our parametrization of $H(t)$ also
gives rise to different evolutions of scale factor that is being studied in
this work.

We shall make note that, $\alpha $ \& $\beta $ are two model parameters and
the dynamics of obtained models (in table-2) or the behavior of cosmological
parameters $a(t),$ $H(t),$ $q(t),$ $\rho (t),$ $w(t)$ all heavily depend on
these two. In the next section we shall discuss the behavior of different
cosmological parameters in view of the positive value of the model
parameters $\alpha $ \& $\beta $ and discuss their analytical bounds. A lot
of studies have been done on model-I ($\Lambda $CDM), model-II (PLC) and
model-III (BM). So, we do not dwell in these known models and try to explore
the other models only i.e. models-IV--XII.

\section{Dynamics of models}

\qquad The expressions for the scale factor, Hubble parameter and
deceleration parameter for model-IV--XII are given in table-2. For the
positive values of $\alpha ,\beta $, their behavior near the singularities
(at $t=0$ and $t\rightarrow \infty $) are obtained as

\begin{center}
\begin{tabular}{lllllllllll}
& {\small Table-3} &  &  &  &  &  &  &  &  &  \\ \hline
\multicolumn{1}{|l}{{\small CP}$\downarrow $} & \multicolumn{1}{|l}{\small %
models} & \multicolumn{1}{|l}{\small IV} & \multicolumn{1}{|l}{\small V} & 
\multicolumn{1}{|l}{\small VI} & \multicolumn{1}{|l}{\small VII} & 
\multicolumn{1}{|l}{\small VIII} & \multicolumn{1}{|l}{\small IX} & 
\multicolumn{1}{|l}{\small X} & \multicolumn{1}{|l}{\small XI} & 
\multicolumn{1}{|l|}{\small XII} \\ \hline
\multicolumn{1}{|l}{${\small a(t)}$} & \multicolumn{1}{|l}{${\small t=0}$} & 
\multicolumn{1}{|l}{${\small C}$} & \multicolumn{1}{|l}{${\small C}$} & 
\multicolumn{1}{|l}{${\small Ce}^{2\beta \sqrt{\alpha }}$} & 
\multicolumn{1}{|l}{${\small C\alpha }^{\beta /2}$} & \multicolumn{1}{|l}{$%
{\small C\alpha }^{-\alpha \beta }$} & \multicolumn{1}{|l}{${\small C\alpha }%
^{\beta /2}$} & \multicolumn{1}{|l}{${\small Ce}^{\beta \sqrt{\alpha }}$} & 
\multicolumn{1}{|l}{${\small 0}$} & \multicolumn{1}{|l|}{${\small 0}$} \\ 
\cline{2-11}
\multicolumn{1}{|l}{} & \multicolumn{1}{|l}{${\small t\rightarrow \infty }$}
& \multicolumn{1}{|l}{${\small \infty }$} & \multicolumn{1}{|l}{${\small Ce}%
^{\frac{\pi \beta }{2\sqrt{\alpha }}}$} & \multicolumn{1}{|l}{${\small %
\infty }$} & \multicolumn{1}{|l}{${\small \infty }$} & \multicolumn{1}{|l}{$%
{\small \infty }$} & \multicolumn{1}{|l}{${\small \infty }$} & 
\multicolumn{1}{|l}{${\small \infty }$} & \multicolumn{1}{|l}{${\small C}$}
& \multicolumn{1}{|l|}{${\small C}$} \\ \hline
\multicolumn{1}{|l}{${\small H(t)}$} & \multicolumn{1}{|l}{${\small t=0}$} & 
\multicolumn{1}{|l}{${\small 0}$} & \multicolumn{1}{|l}{$\frac{\beta }{%
\alpha }$} & \multicolumn{1}{|l}{$\frac{{\small \beta }}{\sqrt{{\small %
\alpha }}}$} & \multicolumn{1}{|l}{$\frac{{\small \beta }}{\sqrt{{\small %
\alpha }}}$} & \multicolumn{1}{|l}{${\small 0}$} & \multicolumn{1}{|l}{$%
{\small 0}$} & \multicolumn{1}{|l}{${\small 0}$} & \multicolumn{1}{|l}{$%
{\small \infty }$} & \multicolumn{1}{|l|}{${\small \infty }$} \\ \cline{2-11}
\multicolumn{1}{|l}{} & \multicolumn{1}{|l}{${\small t\rightarrow \infty }$}
& \multicolumn{1}{|l}{${\small \infty }$} & \multicolumn{1}{|l}{${\small 0}$}
& \multicolumn{1}{|l}{${\small 0}$} & \multicolumn{1}{|l}{${\small 0}$} & 
\multicolumn{1}{|l}{${\small \beta }$} & \multicolumn{1}{|l}{${\small 0}$} & 
\multicolumn{1}{|l}{${\small \beta }$} & \multicolumn{1}{|l}{${\small 0}$} & 
\multicolumn{1}{|l|}{${\small 0}$} \\ \hline
\multicolumn{1}{|l}{${\small q(t)}$} & \multicolumn{1}{|l}{${\small t=0}$} & 
\multicolumn{1}{|l}{${\small -\infty }$} & \multicolumn{1}{|l}{${\small -1}$}
& \multicolumn{1}{|l}{${\small -1+}\frac{1}{2\beta \sqrt{\alpha }}$} & 
\multicolumn{1}{|l}{${\small -1}$} & \multicolumn{1}{|l}{${\small -\infty }$}
& \multicolumn{1}{|l}{${\small -\infty }$} & \multicolumn{1}{|l}{${\small %
-\infty }$} & \multicolumn{1}{|l}{${\small -1+}\frac{\alpha }{\beta }$} & 
\multicolumn{1}{|l|}{${\small -1+}\frac{\alpha }{\beta }$} \\ \cline{2-11}
\multicolumn{1}{|l}{} & \multicolumn{1}{|l}{${\small t\rightarrow \infty }$}
& \multicolumn{1}{|l}{${\small -1}$} & \multicolumn{1}{|l}{${\small +\infty }
$} & \multicolumn{1}{|l}{${\small -1}$} & \multicolumn{1}{|l}{${\small -1+}%
\frac{1}{\beta }$} & \multicolumn{1}{|l}{${\small -1}$} & 
\multicolumn{1}{|l}{${\small -1+}\frac{1}{\beta }$} & \multicolumn{1}{|l}{$%
{\small -1}$} & \multicolumn{1}{|l}{${\small +\infty }$} & 
\multicolumn{1}{|l|}{${\small +\infty }$} \\ \hline
\end{tabular}
\end{center}

From the above table we can see that the models-IV--X are free from initial
singularity and starts with a finite initial radius while models-XI,XII have
big bang origin. As $t\rightarrow \infty $ models-IV,VI,VII,VIII,IX,X
diverge while the scale factor takes finite values in models-V,XI,XII.
Models-IV,VI,VIII,X collapse in near future showing these models have finite
time singularity but the singularity can be delayed by larger (smaller in
case of model-IV) values $\beta $. Similarly, looking at the values of HP
and DP near singularities, we can conclude that in case of
models-IV,VIII,IX,X, the Universe starts with zero velocity and infinite
acceleration. In case of Model-V,VI,VII the Universe starts with finite
velocity and finite acceleration while in models-XI,XII the Universe starts
with infinite velocity and finite acceleration. The rate of initial
velocities and initial accelerations for these models depend upon the choice
of model parameters $\alpha $ \& $\beta $. We can observe that in the
models-V,XI,XII, the Universe ceases as $t\rightarrow \infty $ where the
velocity becomes zero and DP becomes $+\infty $ (no acceleration).

As the observations reveal that the total energy budget of the Universe is
dominated by DE ($\sim 70\%$) and accelerates the expansion of the Universe
while non-relativistic baryonic and cold dark matter dominated the total
energy budget at earlier times, causing the deceleration. The cosmological
deceleration-acceleration transition occurred at some time where $q=0$ (or $%
\ddot{a}=0$). So, at present theorists take interest in modelling the
Universe with phase transition from early deceleration to present
acceleration. These kinds of models are considered as viable models as there
is an obvious provision for the structure formation in the Universe during
the decelerated phase and also they can explain the result of observation of
Type Ia supernovae at present. On the other hand, before the discovery of
late time acceleration, theorists were taking interest in modelling the
Universe with early inflation and late time deceleration of the Universe as
early inflation is necessary to explain the origin of the large scale
structure of the cosmos. So, the \textit{deceleration-acceleration} phase
transition is important in current picture while the \textit{%
acceleration-deceleration} phase transition is important in the very early
Universe. In conclusion, we can say a model which has initial acceleration,
middle deceleration and late-time acceleration scenario can be treated as a
better model that can explain all the phenomena explained by observations.

Out of the twelve models listed in table-2, the DP comes out to be constant
in models-I,II,III where as the DP is time-dependent in models-IV--XII. For $%
\alpha ,\beta >0$, models-IV,VIII,X exhibit eternal acceleration and
models-V,VII,IX,XI,XII show transition from initial acceleration to
deceleration; or may accelerate for ever for certain choice of $\alpha $ and 
$\beta $. Only model-VI shows a phase transition from deceleration to
acceleration. The various cases in view of phase transition are analyzed in
the following table.

\begin{center}
\begin{tabular}{clcl}
{\small Table-4} &  &  &  \\ \hline
\multicolumn{1}{|c}{\small Models} & \multicolumn{1}{|l}{\small Transition
type} & \multicolumn{1}{|c}{{\small Phase transition time} ${\small t}_{tr}$}
& \multicolumn{1}{|l|}{{\small constrain on }${\small \alpha }${\small , }$%
{\small \beta }$} \\ \hline
\multicolumn{1}{|c}{\small IV} & \multicolumn{1}{|l}{\small Ever accelerating%
} & \multicolumn{1}{|c}{---} & \multicolumn{1}{|l|}{${\small \alpha },%
{\small \beta >0}$} \\ \hline
\multicolumn{1}{|c}{\small V} & \multicolumn{1}{|l}{{\small Acceleration}$%
\rightarrow ${\small Deceleration}} & \multicolumn{1}{|c}{$\frac{\beta }{2}$}
& \multicolumn{1}{|l|}{${\small \alpha },{\small \beta >0}$} \\ \hline
\multicolumn{1}{|c}{\small VI} & \multicolumn{1}{|l}{{\small Deceleration}$%
\rightarrow ${\small Acceleration}} & \multicolumn{1}{|c}{$\frac{1}{4\beta
^{2}}{\small -\alpha }$} & \multicolumn{1}{|l|}{${\small \alpha },{\small %
\beta >0,\beta \sqrt{\alpha }<}\frac{1}{2}$} \\ 
\multicolumn{1}{|c}{} & \multicolumn{1}{|l}{\small Ever accelerating} & 
\multicolumn{1}{|c}{---} & \multicolumn{1}{|l|}{${\small \alpha },{\small %
\beta >0,\beta \sqrt{\alpha }>}\frac{1}{2}$} \\ \hline
\multicolumn{1}{|c}{\small VII} & \multicolumn{1}{|l}{{\small Acceleration}$%
\rightarrow ${\small Deceleration}} & \multicolumn{1}{|c}{$\frac{\sqrt{%
\alpha }\beta }{\sqrt{1-\beta ^{2}}}$} & \multicolumn{1}{|l|}{${\small %
\alpha },{\small \beta >0,\beta \sqrt{1+\alpha }>}1$} \\ 
\multicolumn{1}{|c}{} & \multicolumn{1}{|l}{\small Ever accelerating} & 
\multicolumn{1}{|c}{---} & \multicolumn{1}{|l|}{${\small \alpha },{\small %
\beta >0},{\small \beta \sqrt{1+\alpha }<}1$} \\ \hline
\multicolumn{1}{|c}{\small VIII} & \multicolumn{1}{|l}{\small Ever
accelerating} & \multicolumn{1}{|c}{---} & \multicolumn{1}{|l|}{${\small %
\alpha ,\beta >0}$} \\ \hline
\multicolumn{1}{|c}{\small IX} & \multicolumn{1}{|l}{{\small Acceleration}$%
\rightarrow ${\small Deceleration}} & \multicolumn{1}{|c}{$\sqrt{\frac{%
\alpha }{1-\beta }}$} & \multicolumn{1}{|l|}{${\small \alpha ,\beta >0},%
{\small \beta +\alpha >1}$} \\ 
\multicolumn{1}{|c}{} & \multicolumn{1}{|l}{\small Ever accelerating} & 
\multicolumn{1}{|c}{---} & \multicolumn{1}{|l|}{${\small \alpha ,\beta >0},%
{\small \beta +\alpha <1}$} \\ \hline
\multicolumn{1}{|c}{\small X} & \multicolumn{1}{|l}{\small Ever accelerating}
& \multicolumn{1}{|c}{---} & \multicolumn{1}{|l|}{${\small \alpha ,\beta >0}$%
} \\ \hline
\multicolumn{1}{|c}{\small XI} & \multicolumn{1}{|l}{{\small Acceleration}$%
\rightarrow ${\small Deceleration}} & \multicolumn{1}{|c}{$\frac{\beta
-\alpha }{2}$} & \multicolumn{1}{|l|}{${\small \alpha ,\beta >0,\beta
>\alpha }$} \\ 
\multicolumn{1}{|c}{} & \multicolumn{1}{|l}{\small Ever Decelerating} & 
\multicolumn{1}{|c}{---} & \multicolumn{1}{|l|}{${\small \alpha ,\beta
>0,\beta <\alpha }$} \\ \hline
\multicolumn{1}{|c}{\small XII} & \multicolumn{1}{|l}{{\small Acceleration}$%
\rightarrow ${\small Deceleration}} & \multicolumn{1}{|c}{$\sqrt{\frac{\beta
-\alpha }{3}}$} & \multicolumn{1}{|l|}{${\small \alpha ,\beta >0,\beta
>\alpha }$} \\ 
\multicolumn{1}{|c}{} & \multicolumn{1}{|l}{\small Ever Decelerating} & 
\multicolumn{1}{|c}{---} & \multicolumn{1}{|l|}{${\small \alpha ,\beta
<0,\beta <\alpha }$} \\ \hline
\end{tabular}
\end{center}

Observations suggest the present value of DP is somewhere in the
neighborhood of $-0.55$. So, to have a better understanding of evolutions of
DP over time for all the models-IV--XII, we plot them (see figure-1) by
choosing the values of $\alpha $ and $\beta $ appropriately such that the
present value of DP $q_{0}$ will be in the neighborhood $-0.55$ or with a
very small positive value of $q_{0}$ (for decelerating models). It may be
noted that the observations favour accelerating models but the decelerating
models are also in agreement with these observations \cite{vishwa-metalic}.
The decelerating models also show nice fit to some data even with zero
cosmological constant if one considers the extinction of light by the
metallic dust ejected from the supernovae explosions \cite{vishwa-metalic}.
With some independent analysis, we have chosen the values of $\alpha $ and $%
\beta $ in model-VI such that $q_{0}\approx -0.55$. This gives the phase
transition time from deceleration to acceleration is around $t_{tr}\approx 3$%
. The time evolution of $q(t)$ for models-IV--XII are

\begin{figure}[tbph]
\centering
\includegraphics[width=0.32\textwidth]{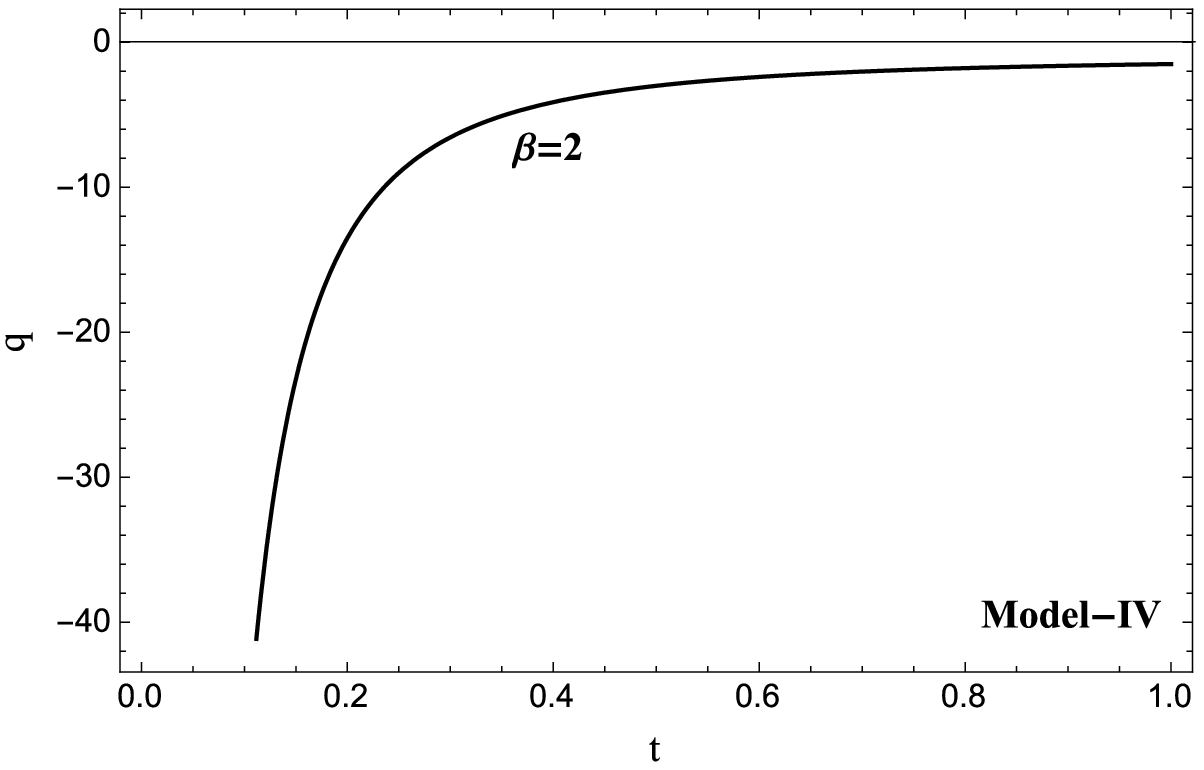} \includegraphics[width=0.32%
\textwidth]{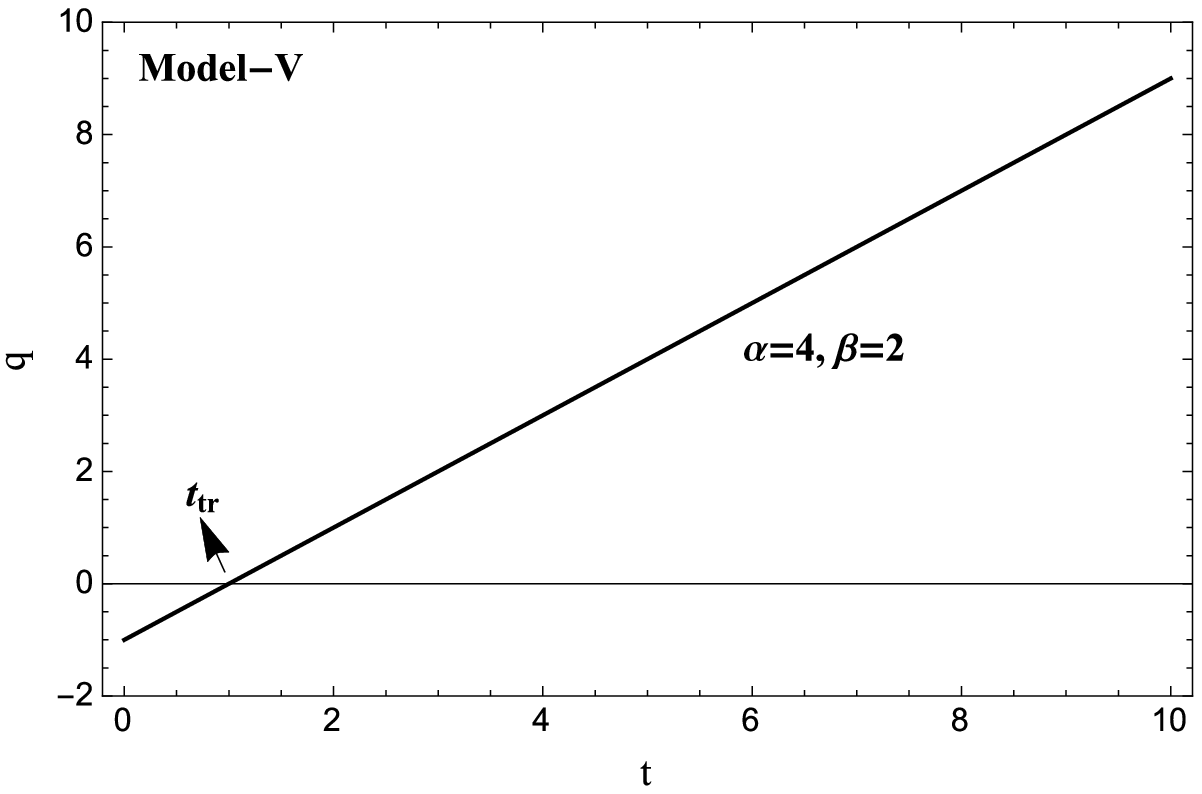} \includegraphics[width=0.32\textwidth]{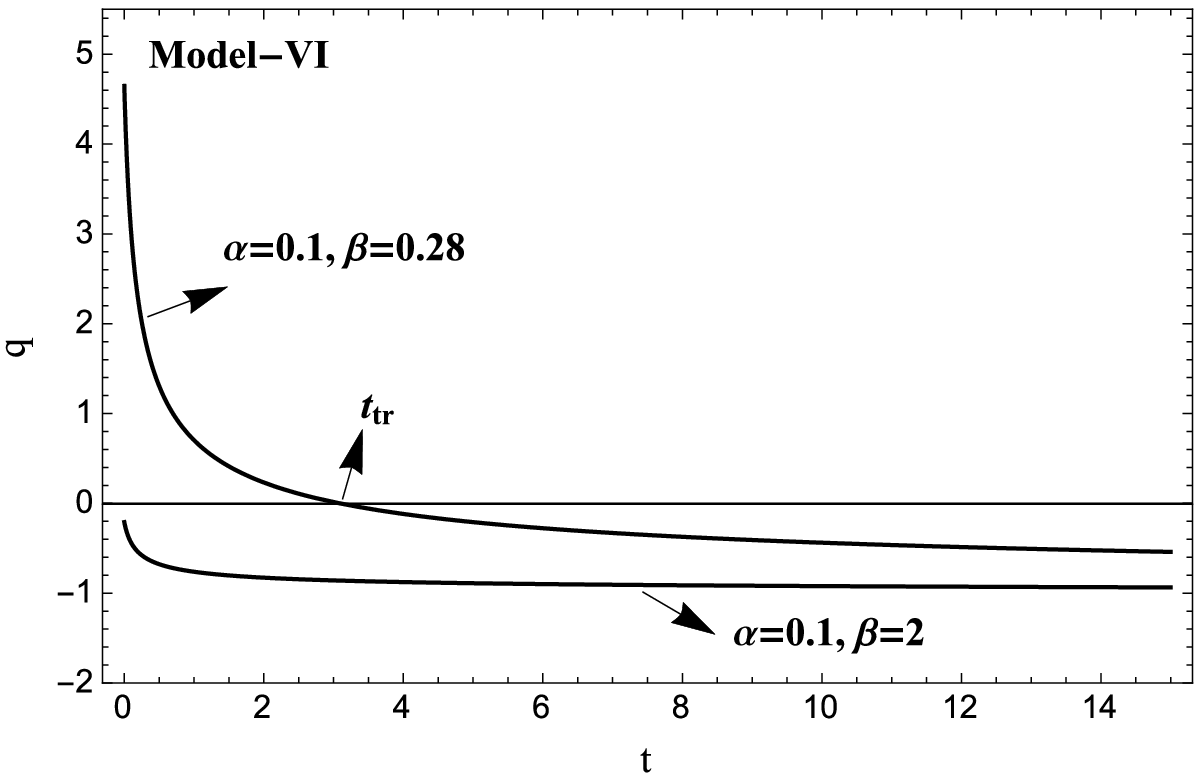} %
\includegraphics[width=0.32\textwidth]{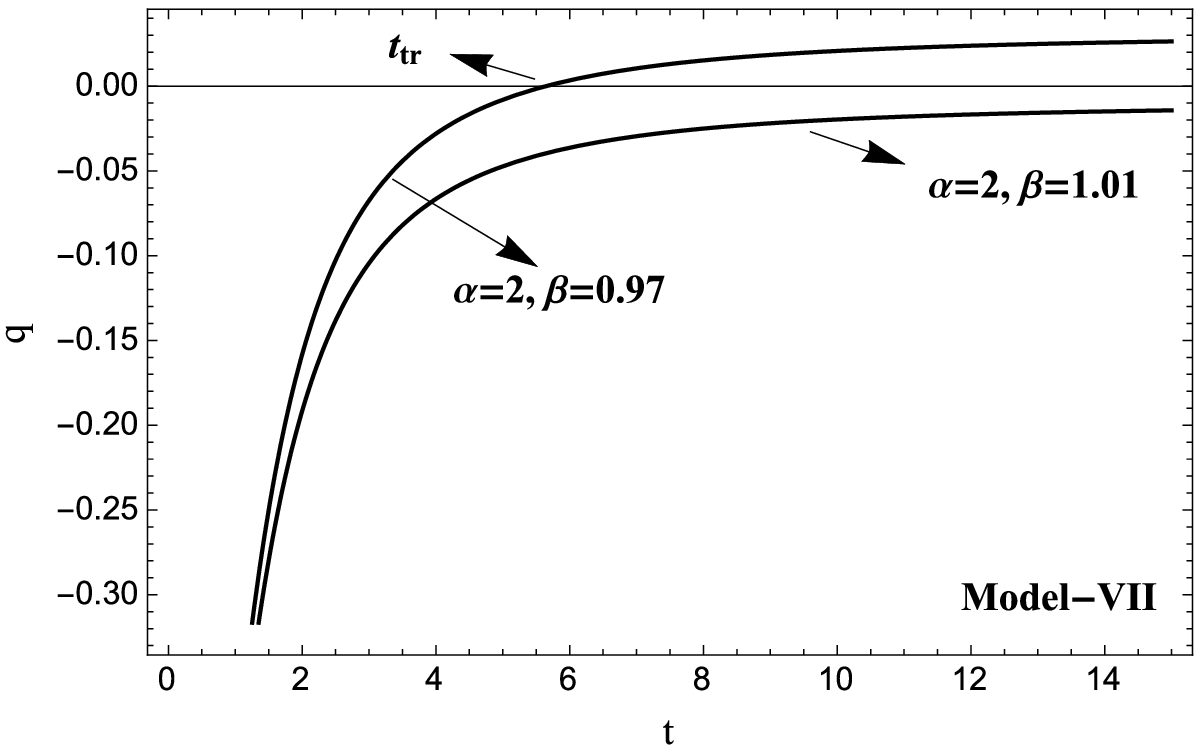} \includegraphics[width=0.32%
\textwidth]{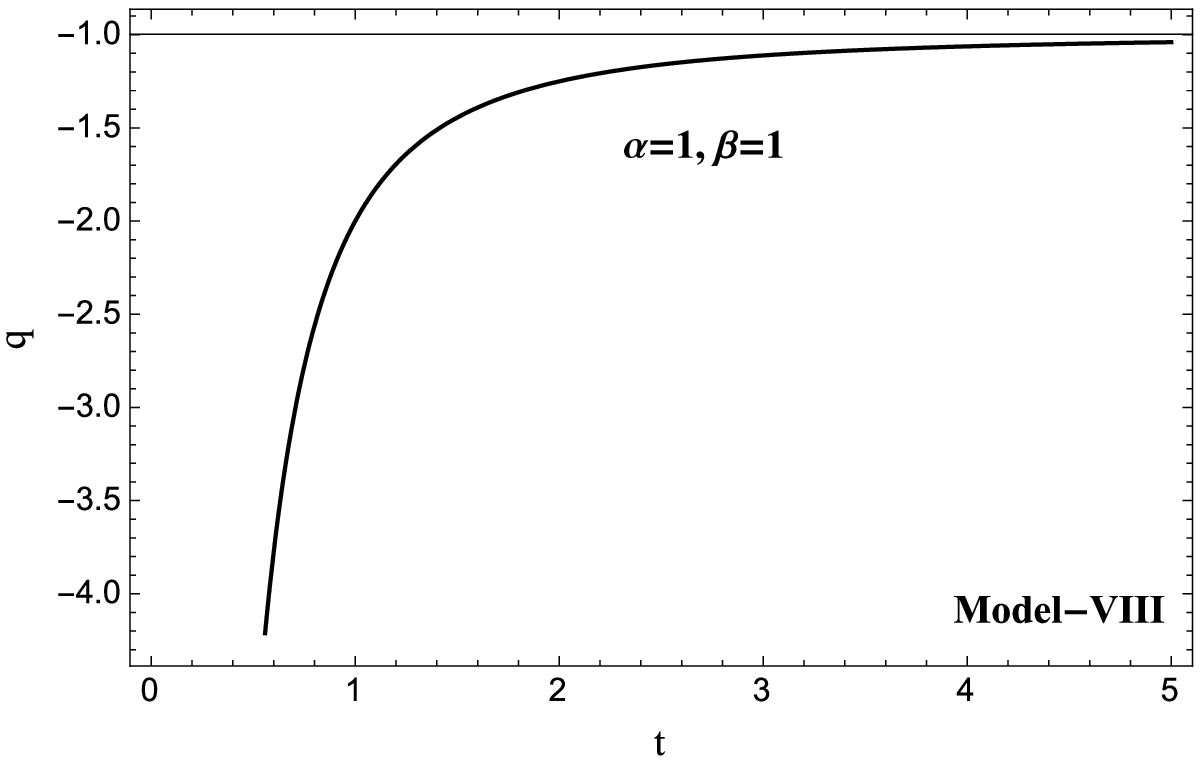} \includegraphics[width=0.32\textwidth]{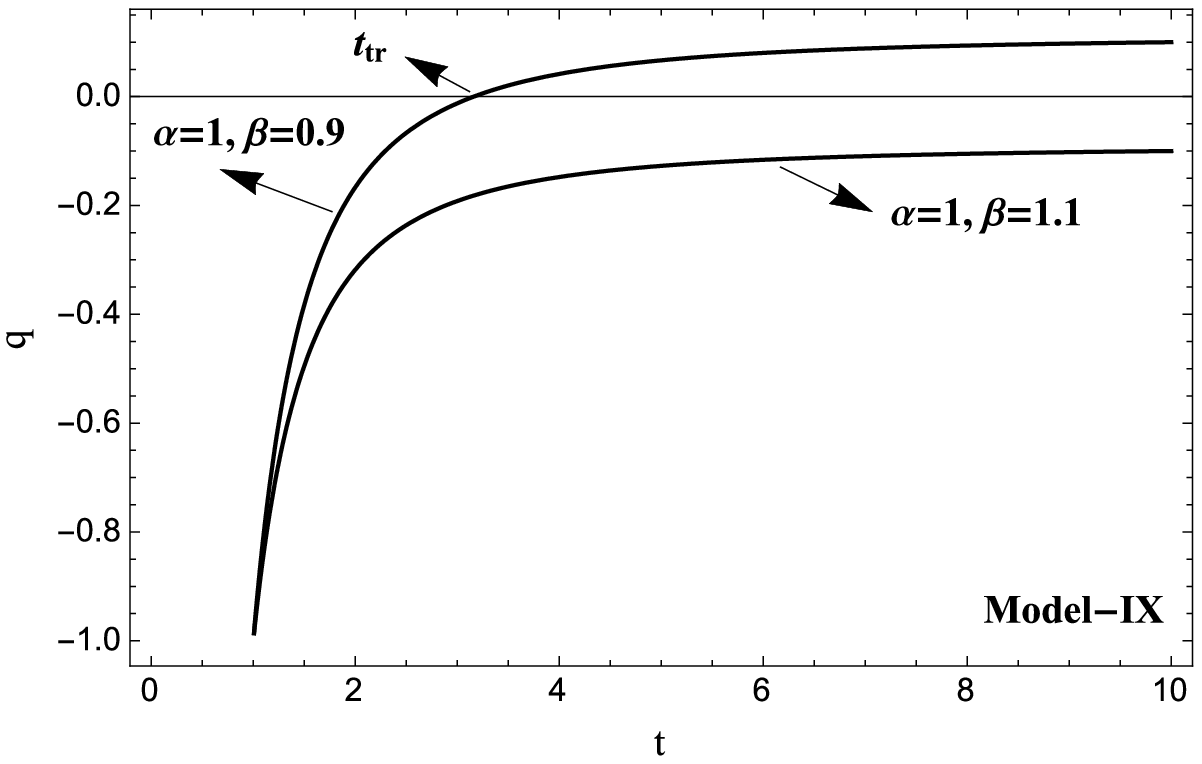} %
\includegraphics[width=0.32\textwidth]{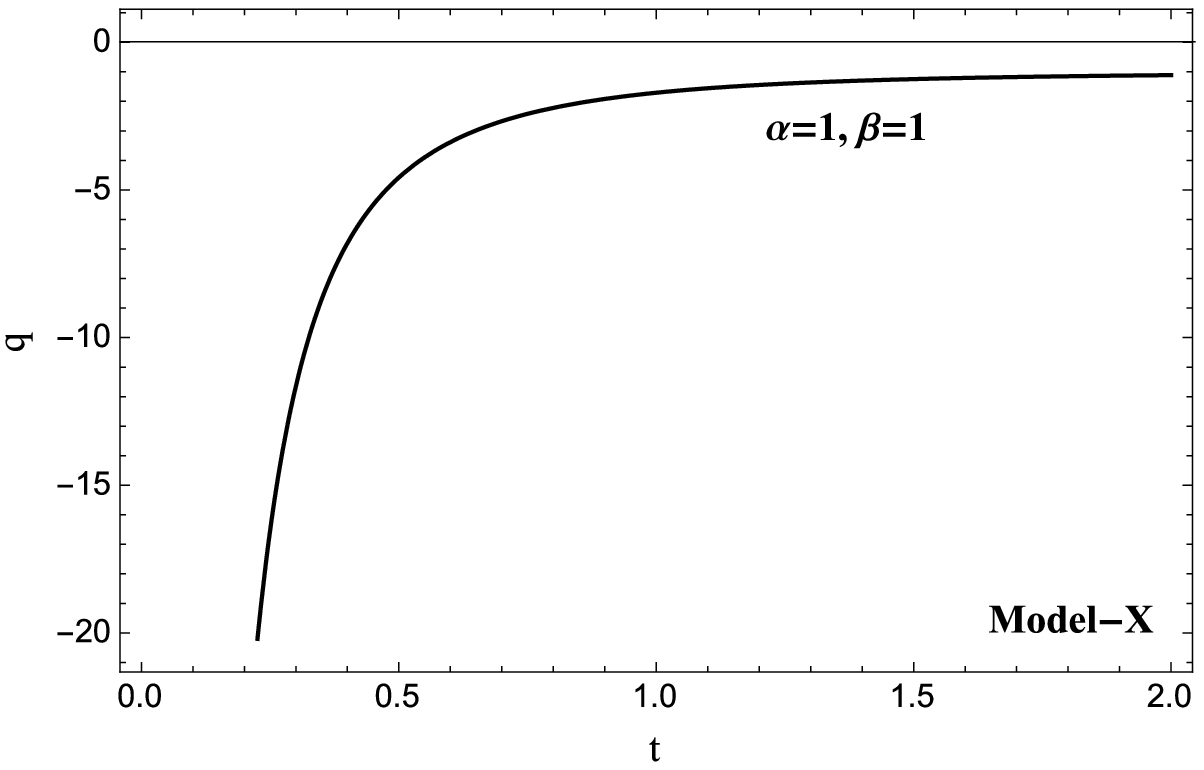} \includegraphics[width=0.32%
\textwidth]{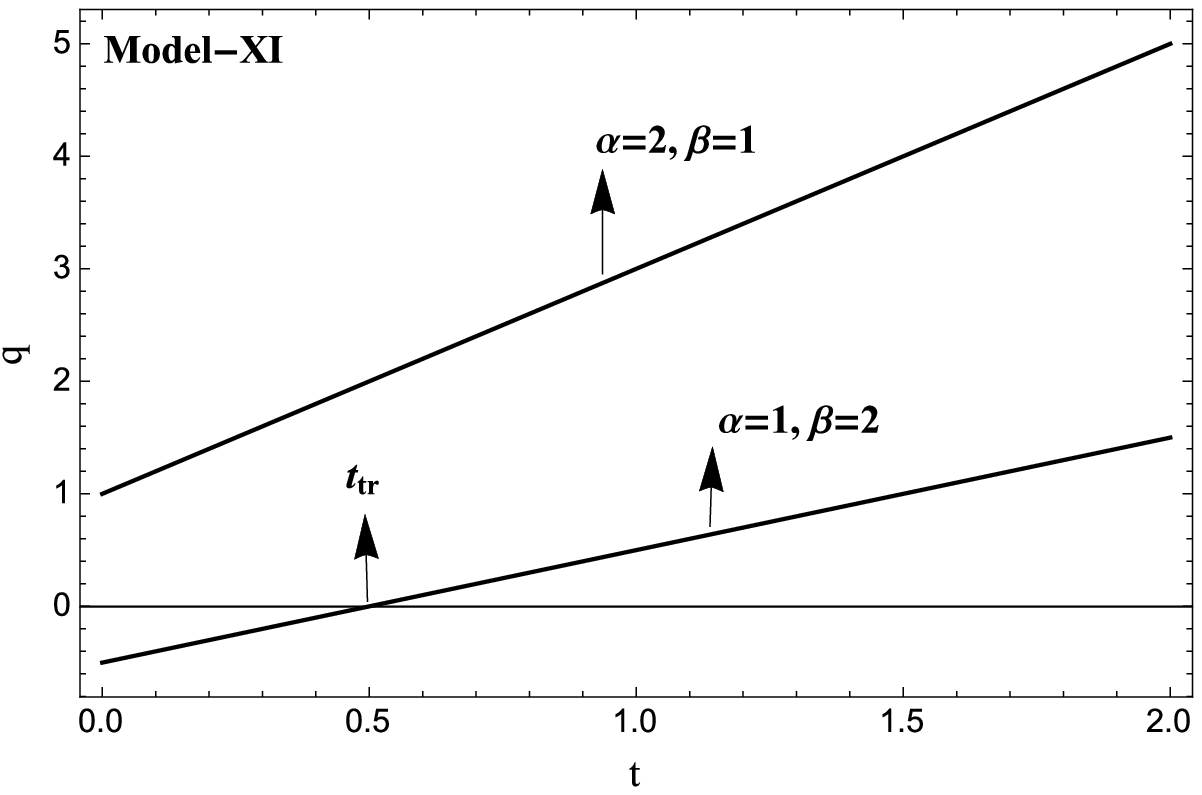} \includegraphics[width=0.32\textwidth]{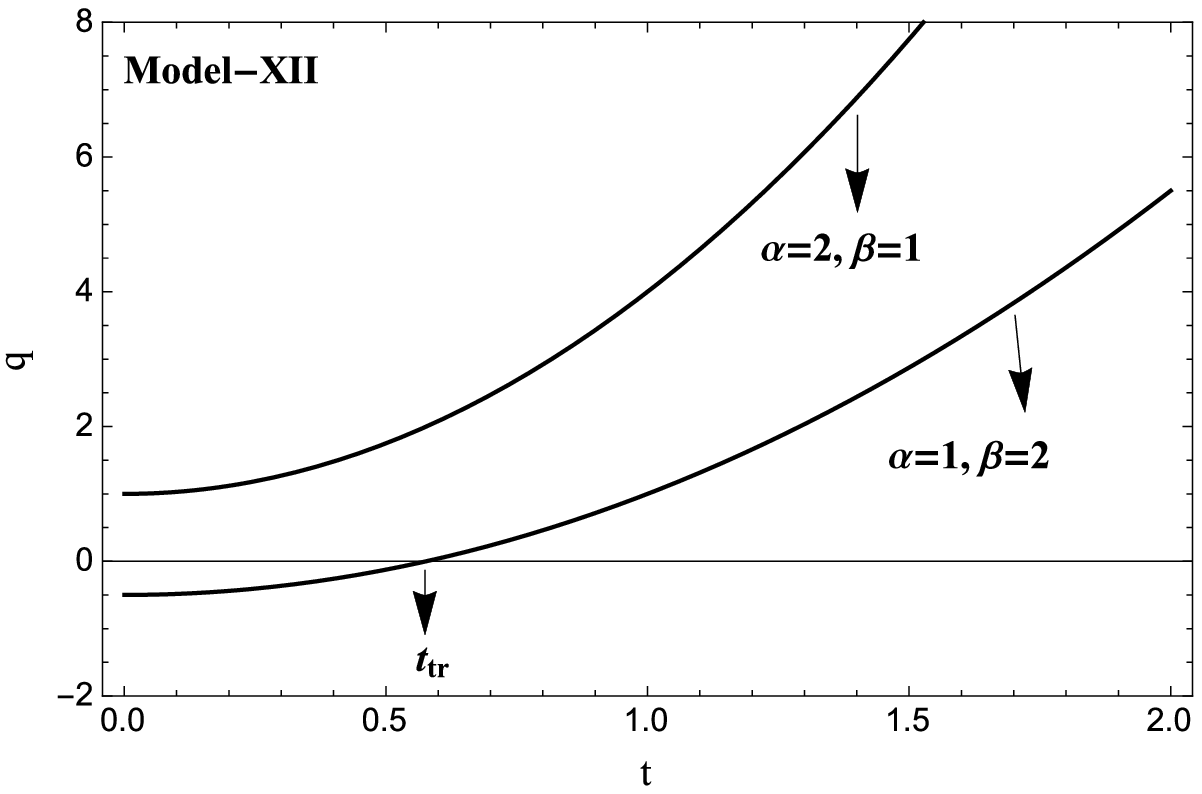}
\caption{Plots for the Deceleration parameter for models-IV--XII. We can see
model-IV (with $\protect\beta =2$), model-VIII (with $\protect\alpha =1,%
\protect\beta =1$), model-X (with $\protect\alpha =1,\protect\beta =1$) show
eternal acceleration while model-V (with $\protect\alpha =4,\protect\beta =2$%
), model-VII (with $\protect\alpha =2,\protect\beta =0.97$), model-IX (with $%
\protect\alpha =1,\protect\beta =0.9$), model-XI (with $\protect\alpha =1,%
\protect\beta =2$), model-XII (with $\protect\alpha =1,\protect\beta =2$)
show transition from acceleration to deceleration. Only model-VI (with $%
\protect\alpha =0.1,\protect\beta =0.28$) shows transition from deceleration
to acceleration. We can also see that the model-VI (with $\protect\alpha %
=0.1,\protect\beta =2$), model-VII (with $\protect\alpha =2,\protect\beta %
=1.01$), model-IX (with $\protect\alpha =1,\protect\beta =1.1$) show eternal
acceleration and model-XI (with $\protect\alpha =2,\protect\beta =1$),
model-XII (with $\protect\alpha =2,\protect\beta =1$) show eternal
deceleration as discussed in table-4. For the plots we have chosen the units
suitably e.g. In case of model-IV, the time axis is scaled such that $0.1$
time unit $=1$ billion years while in model-V, the time axis is scaled such
that $1$ time unit $=1$ billion years and so on.}
\label{fig:q-XII}
\end{figure}

\newpage \qquad The other cosmological parameters $\rho (t),$ $\rho
_{eff}(t),$ $w_{eff}(t)$ for models-IV--XII are obtained as

\begin{tabular}{lll}
\textbf{Model-IV} &  & \textbf{Model-V} \\ 
$\rho (t)=\left[ DC^{-3(1+w)}\right] e^{-3(1+w)\beta \frac{t^{2}}{2}}$ &  & $%
\rho (t)=\left[ DC^{-3(1+w)}\right] e^{-3(1+w)\frac{\beta }{\sqrt{\alpha }}%
\tan ^{-1}\frac{t}{\sqrt{\alpha }}}$ \\ 
$\rho _{eff}(t)=3M_{Pl}^{2}\left[ \beta ^{2}t^{2}+\frac{k}{C^{2}}e^{-\beta
t^{2}}\right] $ &  & $\rho _{eff}(t)=3M_{Pl}^{2}\left[ \frac{\beta ^{2}}{%
\left( t^{2}+\alpha \right) ^{2}}+\frac{k}{C^{2}}e^{-\frac{2\beta }{\sqrt{%
\alpha }}\tan ^{-1}\frac{t}{\sqrt{\alpha }}}\right] $ \\ 
$w_{eff}(t)=\frac{1}{3}\frac{\left( -3-\frac{2}{\beta }\frac{1}{t^{2}}%
\right) \beta ^{2}t^{2}-\frac{k}{C^{2}}e^{-\beta t^{2}}}{\left[ \beta
^{2}t^{2}+\frac{k}{C^{2}}e^{-\beta t^{2}}\right] }$ &  & $w_{eff}(t)=\frac{1%
}{3}\frac{\left( -3+\frac{4}{\beta }t\right) \frac{\beta ^{2}}{\left(
t^{2}+\alpha \right) ^{2}}-\frac{k}{C^{2}}e^{-\frac{2\beta }{\sqrt{\alpha }}%
\tan ^{-1}\frac{t}{\sqrt{\alpha }}}}{\frac{\beta ^{2}}{\left( t^{2}+\alpha
\right) ^{2}}+\frac{k}{C^{2}}e^{-\frac{2\beta }{\sqrt{\alpha }}\tan ^{-1}%
\frac{t}{\sqrt{\alpha }}}}$%
\end{tabular}

\begin{tabular}{lll}
\textbf{Model-VI} &  & \textbf{Model-VII} \\ 
$\rho (t)=\left[ DC^{-3(1+w)}\right] e^{-6(1+w)\beta \sqrt{t+\alpha }}$ &  & 
$\rho (t)=\left[ DC^{-3(1+w)}\right] \left( t+\sqrt{t^{2}+\alpha }\right)
^{-3(1+w)\beta }$ \\ 
$\rho _{eff}(t)=3M_{Pl}^{2}\left[ \frac{\beta ^{2}}{t+\alpha }+\frac{k}{C^{2}%
}e^{-4\beta \sqrt{t+\alpha }}\right] $ &  & $\rho _{eff}(t)=3M_{Pl}^{2}\left[
\frac{\beta ^{2}}{t^{2}+\alpha }+\frac{k}{C^{2}\left( t+\sqrt{t^{2}+\alpha }%
\right) ^{2\beta }}\right] $ \\ 
$w_{eff}(t)=\frac{1}{3}\frac{\left( -3+\frac{1}{\beta }\frac{1}{\sqrt{%
t+\alpha }}\right) \frac{\beta ^{2}}{t+\alpha }-\frac{k}{C^{2}}e^{-4\beta 
\sqrt{t+\alpha }}}{\frac{\beta ^{2}}{t+\alpha }+\frac{k}{C^{2}}e^{-4\beta 
\sqrt{t+\alpha }}}$ &  & $w_{eff}(t)=\frac{1}{3}\frac{\left( -3+\frac{2}{%
\beta }\frac{t}{\sqrt{t^{2}+\alpha }}\right) \frac{\beta ^{2}}{t^{2}+\alpha }%
-\frac{k}{C^{2}\left( t+\sqrt{t^{2}+\alpha }\right) ^{2\beta }}}{\frac{\beta
^{2}}{t^{2}+\alpha }+\frac{k}{C^{2}\left( t+\sqrt{t^{2}+\alpha }\right)
^{2\beta }}}$%
\end{tabular}

\begin{tabular}{lll}
\textbf{Model-VIII} &  & \textbf{Model-IX} \\ 
$\rho (t)=\left[ DC^{-3(1+w)}\right] e^{-3(1+w)\beta t}\left( t+\alpha
\right) ^{3(1+w)\alpha \beta }$ &  & $\rho (t)=\left[ DC^{-3(1+w)}\right]
\left( t^{2}+\alpha \right) ^{-3(1+w)\frac{\beta }{2}}$ \\ 
$\rho _{eff}(t)=3M_{Pl}^{2}\left[ \frac{\beta ^{2}t^{2}}{\left( t+\alpha
\right) ^{2}}+\frac{k}{C^{2}e^{2\beta t}\left( t+\alpha \right) ^{-2\alpha
\beta }}\right] $ &  & $\rho _{eff}(t)=3M_{Pl}^{2}\left[ \frac{\beta
^{2}t^{2}}{\left( t^{2}+\alpha \right) ^{2}}+\frac{k}{C^{2}\left(
t^{2}+\alpha \right) ^{\beta }}\right] $ \\ 
$w_{eff}(t)=\frac{1}{3}\frac{\left( -3-\frac{2\alpha }{\beta }\frac{1}{t^{2}}%
\right) \frac{\beta ^{2}t^{2}}{\left( t+\alpha \right) ^{2}}-\frac{k}{%
C^{2}e^{2\beta t}\left( t+\alpha \right) ^{-2\alpha \beta }}}{\frac{\beta
^{2}t^{2}}{\left( t+\alpha \right) ^{2}}+\frac{k}{C^{2}e^{2\beta t}\left(
t+\alpha \right) ^{-2\alpha \beta }}}$ &  & $w_{eff}(t)=\frac{1}{3}\frac{%
\left( -3+\frac{2}{\beta }-\frac{2\alpha }{\beta }\frac{1}{t^{2}}\right) 
\frac{\beta ^{2}t^{2}}{\left( t^{2}+\alpha \right) ^{2}}-\frac{k}{%
C^{2}\left( t^{2}+\alpha \right) ^{\beta }}}{\frac{\beta ^{2}t^{2}}{\left(
t^{2}+\alpha \right) ^{2}}+\frac{k}{C^{2}\left( t^{2}+\alpha \right) ^{\beta
}}}$%
\end{tabular}

\begin{tabular}{lll}
\textbf{Model-X} &  & \textbf{Model-XI} \\ 
$\rho (t)=\left[ DC^{-3(1+w)}\right] e^{-3(1+w)\beta \sqrt{t^{2}+\alpha }}$
&  & $\rho (t)=\left[ DC^{-3(1+w)}\right] \left( \frac{t}{t+\alpha }\right)
^{-3(1+w)\frac{\beta }{\alpha }}$ \\ 
$\rho _{eff}(t)=3M_{Pl}^{2}\left[ \frac{\beta ^{2}t^{2}}{\left( t^{2}+\alpha
\right) }+\frac{k}{C^{2}e^{2\beta \sqrt{t^{2}+\alpha }}}\right] $ &  & $\rho
_{eff}(t)=3M_{Pl}^{2}\left[ \frac{\beta ^{2}}{t^{2}\left( t+\alpha \right)
^{2}}+\frac{k}{C^{2}\left( \frac{t}{t+\alpha }\right) ^{2\frac{\beta }{%
\alpha }}}\right] $ \\ 
$w_{eff}(t)=\frac{1}{3}\frac{\left( -3-\frac{2\alpha }{\beta }\frac{1}{t^{2}%
\sqrt{t^{2}+\alpha }}\right) \frac{\beta ^{2}t^{2}}{t^{2}+\alpha }-\frac{k}{%
C^{2}e^{2\beta \sqrt{t^{2}+\alpha }}}}{\frac{\beta ^{2}t^{2}}{t^{2}+\alpha }+%
\frac{k}{C^{2}e^{2\beta \sqrt{t^{2}+\alpha }}}}$ &  & $w_{eff}(t)=\frac{1}{3}%
\frac{\left( -3+\frac{2\alpha }{\beta }+\frac{2}{\beta }t\right) \frac{\beta
^{2}}{t^{2}\left( t+\alpha \right) ^{2}}-\frac{k}{C^{2}\left( \frac{t}{%
t+\alpha }\right) ^{2\frac{\beta }{\alpha }}}}{\frac{\beta ^{2}}{t^{2}\left(
t+\alpha \right) ^{2}}+\frac{k}{C^{2}\left( \frac{t}{t+\alpha }\right) ^{2%
\frac{\beta }{\alpha }}}}$%
\end{tabular}

\begin{tabular}{lll}
\textbf{Model-XII} &  & \ \ \ \ \ \ \ \ \ \ \ \ \ \ \ \ \ \ \ \ \ \ \ \ \ \
\ \ \ \ \ \ \ \ \ \ \ \ \ \ \ \ \ \ \ \ \ \ \ \ \ \ \ \ \  \\ 
$\rho (t)=\left[ DC^{-3(1+w)}\right] \left( \frac{t^{2}}{t^{2}+\alpha }%
\right) ^{-3(1+w)\frac{\beta }{2\alpha }}$ &  &  \\ 
$\rho _{eff}(t)=3M_{Pl}^{2}\left[ \frac{\beta ^{2}}{t^{2}\left( t^{2}+\alpha
\right) ^{2}}+\frac{k}{C^{2}\left( \frac{t^{2}}{t^{2}+\alpha }\right) ^{%
\frac{\beta }{\alpha }}}\right] $ &  &  \\ 
$w_{eff}(t)=\frac{1}{3}\frac{\left( -3+\frac{2\alpha }{\beta }+\frac{6}{%
\beta }t^{2}\right) \frac{\beta ^{2}}{t^{2}\left( t^{2}+\alpha \right) ^{2}}-%
\frac{k}{C^{2}\left( \frac{t^{2}}{t^{2}+\alpha }\right) ^{\frac{\beta }{%
\alpha }}}}{\frac{\beta ^{2}}{t^{2}\left( t^{2}+\alpha \right) ^{2}}+\frac{k%
}{C^{2}\left( \frac{t^{2}}{t^{2}+\alpha }\right) ^{\frac{\beta }{\alpha }}}}$
&  & 
\end{tabular}

\subsection{Negative $\protect\beta $ consideration}

\qquad We discuss the possibility of taking negative value of $\beta $
together with negative $\alpha $ in certain models giving rise to some new
cosmologies. In this work, we consider negative $\beta $ together with
negative $\alpha $ in models-XI,XII only. In other models one can work out
for negative $\alpha ,\beta $ in models-III,V,VIII,IX where $\alpha >t$.
This kind of analysis have been done by Nojiri and Odintsov \cite{H-odin1}, 
\cite{H-odin2}, \cite{H-odin3} to study the future finite time singularity
where they have taken $\alpha =t_{s}\rightarrow $future singularity time.

So, for negative $\alpha ,\beta $ we obtain the cosmological parameters for
models-XI,XII as

\begin{tabular}{lll}
\textbf{Model-XI}$^{\mathbf{\ast }}$ &  & \textbf{Model-XII}$^{\mathbf{\ast }%
}$ \\ 
$H(t)=\frac{\beta }{t\left( \alpha -t\right) }$ &  & $H(t)=\frac{\beta }{%
t\left( \alpha -t^{2}\right) }$ \\ 
$a(t)=C\left( \frac{t}{\alpha -t}\right) ^{\frac{\beta }{\alpha }}$ &  & $%
a(t)=C\left( \frac{t^{2}}{\alpha -t^{2}}\right) ^{\frac{\beta }{2\alpha }}$
\\ 
$q(t)=-1+\frac{\alpha }{\beta }-\frac{2}{\beta }t$ &  & $q(t)=-1+\frac{%
\alpha }{\beta }-\frac{3}{\beta }t^{2}$ \\ 
$\rho (t)=\left[ DC^{-3(1+w)}\right] \left( \frac{t}{\alpha -t}\right)
^{-3(1+w)\frac{\beta }{\alpha }}$ &  & $\rho (t)=\left[ DC^{-3(1+w)}\right]
\left( \frac{t^{2}}{\alpha -t^{2}}\right) ^{-3(1+w)\frac{\beta }{2\alpha }}$
\\ 
$\rho _{eff}(t)=3M_{Pl}^{2}\left[ \frac{\beta ^{2}}{t^{2}\left( \alpha
-t\right) ^{2}}+\frac{k}{C^{2}\left( \frac{t}{\alpha -t}\right) ^{2\frac{%
\beta }{\alpha }}}\right] $ &  & $\rho _{eff}(t)=3M_{Pl}^{2}\left[ \frac{%
\beta ^{2}}{t^{2}\left( \alpha -t^{2}\right) ^{2}}+\frac{k}{C^{2}\left( 
\frac{t^{2}}{\alpha -t^{2}}\right) ^{\frac{\beta }{\alpha }}}\right] $ \\ 
$w_{eff}(t)=\frac{1}{3}\frac{\left( -3+\frac{2\alpha }{\beta }-\frac{4}{%
\beta }t\right) \frac{\beta ^{2}}{t^{2}\left( \alpha -t\right) ^{2}}-\frac{k%
}{C^{2}\left( \frac{t}{\alpha -t}\right) ^{2\frac{\beta }{\alpha }}}}{\frac{%
\beta ^{2}}{t^{2}\left( \alpha -t\right) ^{2}}+\frac{k}{C^{2}\left( \frac{t}{%
\alpha -t}\right) ^{2\frac{\beta }{\alpha }}}}$ &  & $w_{eff}(t)=\frac{1}{3}%
\frac{\left( -3+\frac{2\alpha }{\beta }-\frac{6}{\beta }t^{2}\right) \frac{%
\beta ^{2}}{t^{2}\left( \alpha -t^{2}\right) ^{2}}-\frac{k}{C^{2}\left( 
\frac{t^{2}}{\alpha -t^{2}}\right) ^{\frac{\beta }{\alpha }}}}{\frac{\beta
^{2}}{t^{2}\left( \alpha -t^{2}\right) ^{2}}+\frac{k}{C^{2}\left( \frac{t^{2}%
}{\alpha -t^{2}}\right) ^{\frac{\beta }{\alpha }}}}$%
\end{tabular}

For both these models, Hubble parameter and scale factor both diverge in
finite time and show big rip singularity in near future.

\subsection{Observational constrain for models showing DEC$\rightarrow $ACC
transition}

\qquad To constrain the model parameters $\alpha $ and $\beta $ and to
compare our results with observation, we also re-write the DP and HP that
are given as functions of cosmic time $t$, in terms of redshift $z$ ($=\frac{%
a_{0}}{a}-1$, where $a_{0}$ is the value of scale factor at present time $%
t=t_{0}$) using the relation between $t$ and $z$ for the models with
deceleration$\rightarrow $acceleration transition i.e. for models-VI,XI$%
^{\ast }$,XII$^{\ast }$.

\textbf{Model-VI}

$t(z)=-\alpha +\left[ \sqrt{t_{0}+\alpha }-\frac{1}{2\beta }\ln (1+z)\right]
^{2}$

$q(z)=-1+\frac{1}{\beta }\left[ \sqrt{t_{0}+\alpha }-\frac{1}{2\beta }\ln
(1+z)\right] ^{-1}$

$H(z)=H_{0}\left[ 1-\frac{\ln (1+z)}{2\beta \sqrt{t_{0}+\alpha }}\right]
^{-1}$

\textbf{Model-XI}$^{\ast }$

$t(z)=\alpha \left[ 1+\left( \frac{\alpha }{t_{0}}-1\right) \left(
1+z\right) ^{\frac{\alpha }{\beta }}\right] ^{-1}$

$q(z)=-1+\frac{\alpha }{\beta }-\frac{2\alpha }{\beta }\left[ 1+\left( \frac{%
\alpha }{t_{0}}-1\right) \left( 1+z\right) ^{\frac{\alpha }{\beta }}\right]
^{-1}$

$H(z)=\frac{H_{0}t_{0}^{2}}{\alpha ^{2}}\left\{ \left( 1+z\right) ^{-\frac{%
\alpha }{2\beta }}+\left( \frac{\alpha }{t_{0}}-1\right) \left( 1+z\right) ^{%
\frac{\alpha }{2\beta }}\right\} ^{2}$

\textbf{Model-XII}$^{\ast }$

$t(z)=\sqrt{\alpha }\left[ 1+\left( \frac{\alpha }{t_{0}^{2}}-1\right)
\left( 1+z\right) ^{\frac{2\alpha }{\beta }}\right] ^{-\frac{1}{2}}$

$q(z)=-1+\frac{\alpha }{\beta }-\frac{3\alpha }{\beta }\left[ 1+\left( \frac{%
\alpha }{t_{0}^{2}}-1\right) \left( 1+z\right) ^{\frac{2\alpha }{\beta }}%
\right] ^{-1}$

$H(z)=\frac{H_{0}t_{0}^{3}}{\alpha ^{\frac{3}{2}}}\left\{ \left( 1+z\right)
^{-\frac{4\alpha }{3\beta }}+\left( \frac{\alpha }{t_{0}^{2}}-1\right)
\left( 1+z\right) ^{\frac{2\alpha }{3\beta }}\right\} ^{\frac{3}{2}}$

We find the observational constraints on both of the model parameters $%
\alpha $ and $\beta $ to the latest $28$ data points of $H(z)$ in the
redshift range $0.100\leqslant z\leqslant 2.3$ (see table-5). The
observational data consist of measurements of the Hubble parameter at
particular redshifts with the corresponding standard deviations ($\sigma
_{H} $) given by

\begin{center}
\begin{tabular}{c|c|c|c|c|c|c|c|}
{\small Table-5} & \multicolumn{7}{c|}{\small Hubble parameter vs redshift
data} \\ \hline
\multicolumn{1}{|c|}{${\small z}$} & ${\small H(z)}$ $\left( \frac{kms^{-1}}{%
Mpc}\right) $ & ${\small \sigma }_{H}$ $\left( \frac{kms^{-1}}{Mpc}\right) $
& ${\small Ref.}$ & ${\small z}$ & ${\small H(z)}$ $\left( \frac{kms^{-1}}{%
Mpc}\right) $ & ${\small \sigma }_{H}$ $\left( \frac{kms^{-1}}{Mpc}\right) $
& ${\small Ref.}$ \\ \hline
\multicolumn{1}{|c|}{${\small 0.100}$} & ${\small 69}$ & ${\small 12}$ & 
{\small \cite{Simon}} & ${\small 0.730}$ & ${\small 97.3}$ & ${\small 7}$ & 
{\small \cite{Blake}} \\ \hline
\multicolumn{1}{|c|}{${\small 0.170}$} & ${\small 83}$ & ${\small 8}$ & 
\multicolumn{1}{|c|}{{\small \cite{Simon}}} & ${\small 0.781}$ & ${\small 105%
}$ & ${\small 12}$ & {\small \cite{Moresco}} \\ \hline
\multicolumn{1}{|c|}{${\small 0.179}$} & ${\small 75}$ & ${\small 4}$ & 
\multicolumn{1}{|c|}{{\small \cite{Moresco}}} & ${\small 0.875}$ & ${\small %
125}$ & ${\small 17}$ & {\small \cite{Moresco}} \\ \hline
\multicolumn{1}{|c|}{${\small 0.199}$} & ${\small 75}$ & ${\small 5}$ & 
\multicolumn{1}{|c|}{{\small \cite{Moresco}}} & ${\small 0.880}$ & ${\small %
90}$ & ${\small 40}$ & {\small \cite{Stern}} \\ \hline
\multicolumn{1}{|c|}{${\small 0.270}$} & ${\small 77}$ & ${\small 14}$ & 
\multicolumn{1}{|c|}{{\small \cite{Simon}}} & ${\small 0.900}$ & ${\small 117%
}$ & ${\small 23}$ & {\small \cite{Simon}} \\ \hline
\multicolumn{1}{|c|}{${\small 0.320}$} & ${\small 79.2}$ & ${\small 5.6}$ & 
\multicolumn{1}{|c|}{{\small \cite{Cuesta}}} & ${\small 1.037}$ & ${\small %
154}$ & ${\small 20}$ & {\small \cite{Moresco}} \\ \hline
\multicolumn{1}{|c|}{${\small 0.352}$} & ${\small 83}$ & ${\small 14}$ & 
\multicolumn{1}{|c|}{{\small \cite{Moresco}}} & ${\small 1.300}$ & ${\small %
168}$ & ${\small 17}$ & {\small \cite{Simon}} \\ \hline
\multicolumn{1}{|c|}{${\small 0.400}$} & ${\small 95}$ & ${\small 17}$ & 
\multicolumn{1}{|c|}{{\small \cite{Simon}}} & ${\small 1.363}$ & ${\small 160%
}$ & ${\small 33.6}$ & {\small \cite{Moresco2}} \\ \hline
\multicolumn{1}{|c|}{${\small 0.440}$} & ${\small 82.6}$ & ${\small 7.8}$ & 
\multicolumn{1}{|c|}{{\small \cite{Blake}}} & ${\small 1.430}$ & ${\small 177%
}$ & ${\small 18}$ & {\small \cite{Simon}} \\ \hline
\multicolumn{1}{|c|}{${\small 0.480}$} & ${\small 97}$ & ${\small 62}$ & 
\multicolumn{1}{|c|}{{\small \cite{Stern}}} & ${\small 1.530}$ & ${\small 140%
}$ & ${\small 14}$ & {\small \cite{Simon}} \\ \hline
\multicolumn{1}{|c|}{${\small 0.570}$} & ${\small 100.3}$ & ${\small 3.7}$ & 
\multicolumn{1}{|c|}{{\small \cite{Cuesta}}} & ${\small 1.750}$ & ${\small %
202}$ & ${\small 40}$ & {\small \cite{Simon}} \\ \hline
\multicolumn{1}{|c|}{${\small 0.593}$} & ${\small 104}$ & ${\small 13}$ & 
\multicolumn{1}{|c|}{{\small \cite{Moresco}}} & ${\small 1.965}$ & ${\small %
186.5}$ & ${\small 50.4}$ & {\small \cite{Moresco2}} \\ \hline
\multicolumn{1}{|c|}{${\small 0.600}$} & ${\small 87.9}$ & ${\small 6.1}$ & 
\multicolumn{1}{|c|}{{\small \cite{Blake}}} & ${\small 2.340}$ & ${\small 222%
}$ & ${\small 7}$ & {\small \cite{Delubac}} \\ \hline
\multicolumn{1}{|c|}{${\small 0.680}$} & ${\small 92}$ & ${\small 8}$ & 
\multicolumn{1}{|c|}{{\small \cite{Moresco}}} & ${\small 2.360}$ & ${\small %
226}$ & ${\small 8}$ & {\small \cite{Font}} \\ \hline
\end{tabular}
\end{center}

To complete the data set, we take $H_{0}=67.8$ $Km/s/Mpc$. The mean values
of model parameters $\alpha $ and $\beta $ are determined by minimizing

\begin{equation*}
\chi _{OHD}^{2}(p_{s})=\sum\limits_{i=1}^{28}\frac{%
[H_{th}(p_{s};z_{i})-H_{obs}(z_{i})]^{2}}{\sigma _{H(z_{i})}^{2}}
\end{equation*}%
where $p_{s}$ denotes the parameters of the model, $H_{th}$ is the
theoretical (model based) value for the Hubble parameter, $H_{obs}$ is the
observed one, $\sigma _{H(z_{i})}$ is the standard error in the observed
value, and the summation runs over 28 observational data points at redshifts 
$z_{i}$.

From our analysis, the model-VI show a poor fit for higher redshifts (not
shown), but models-XI$^{\ast }$ \& XII$^{\ast }$ show nice fit to the Hubble
data compared with $\Lambda CDM$ model and are shown in figure-2 . The
likelihood contours in the $\alpha -\beta $ plane with $1\sigma $ and $%
2\sigma $ error are also obtained for these models and are shown in
figure-3. 
\begin{figure}[tbph]
\centering
\includegraphics[width=0.45\textwidth]{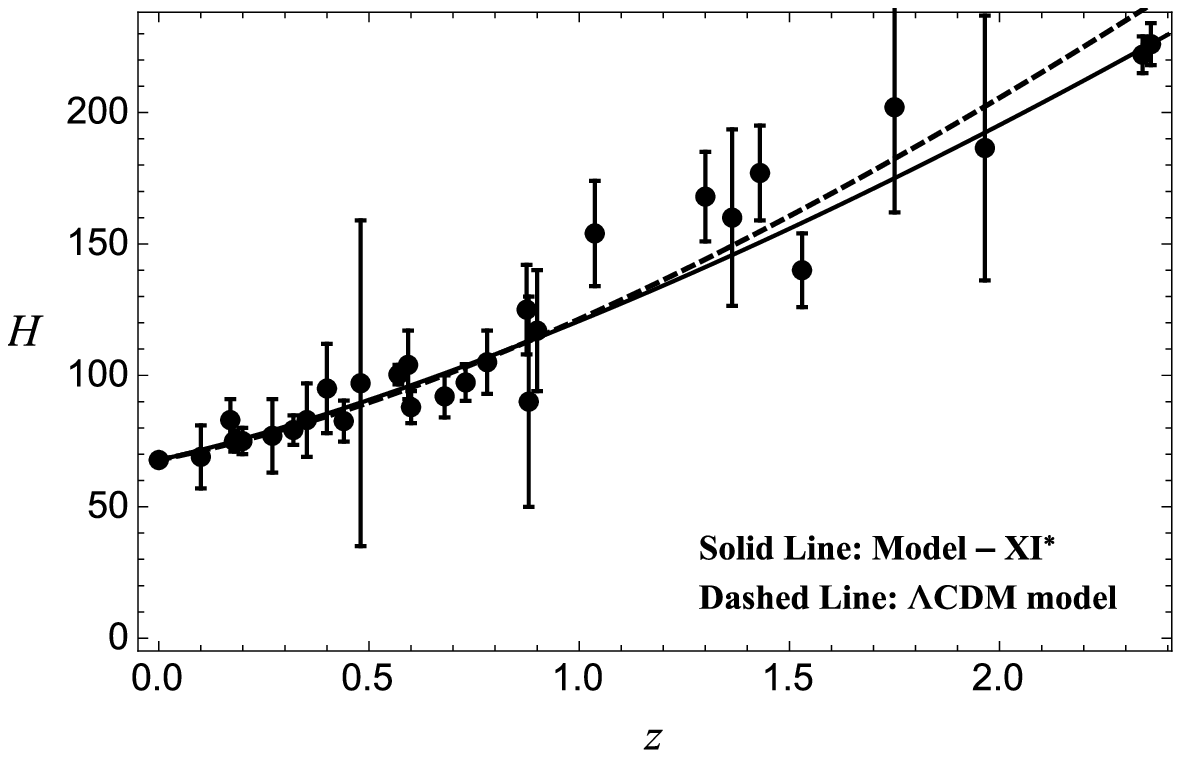} %
\includegraphics[width=0.45\textwidth]{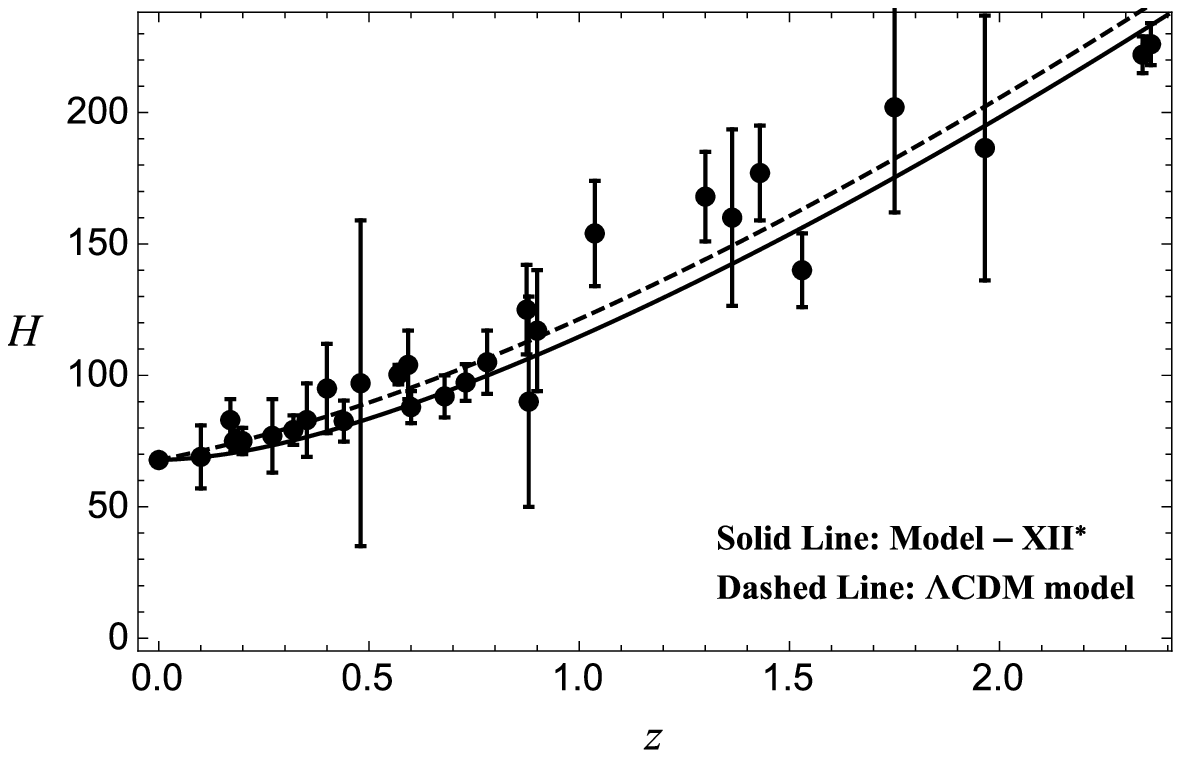}
\caption{This figure corresponds to the latest $H(z)$ data with error bars.
In both the plots, solid lines corresponds to the best fitted behavior for
model-XI$^{\ast }$ \& XII$^{\ast }$ and dotted lines corresponds to $\Lambda
CDM$ model. $H(z)$ is expressed in unit of $Km/s/Mpc$.}
\label{fig:Hz-starXII}
\end{figure}

\begin{figure}[tbph]
\centering
\includegraphics[width=0.39\textwidth]{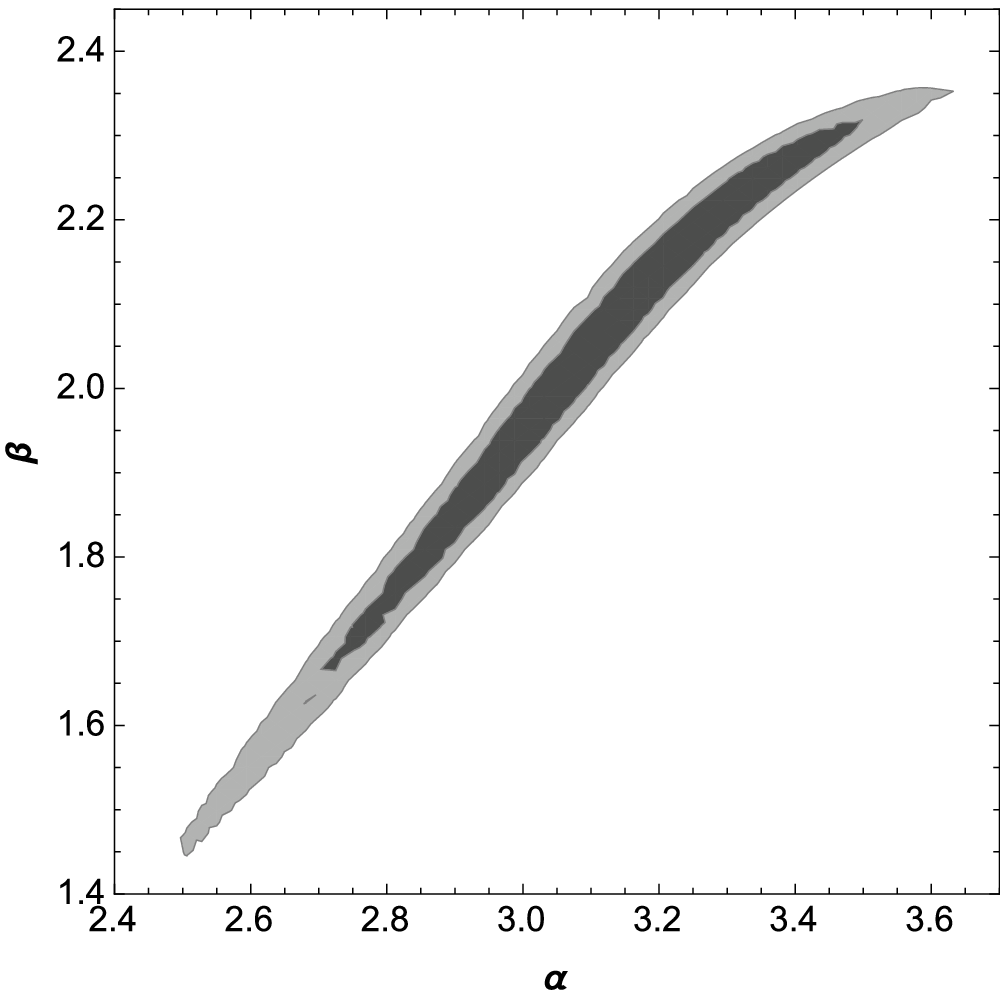} %
\includegraphics[width=0.39\textwidth]{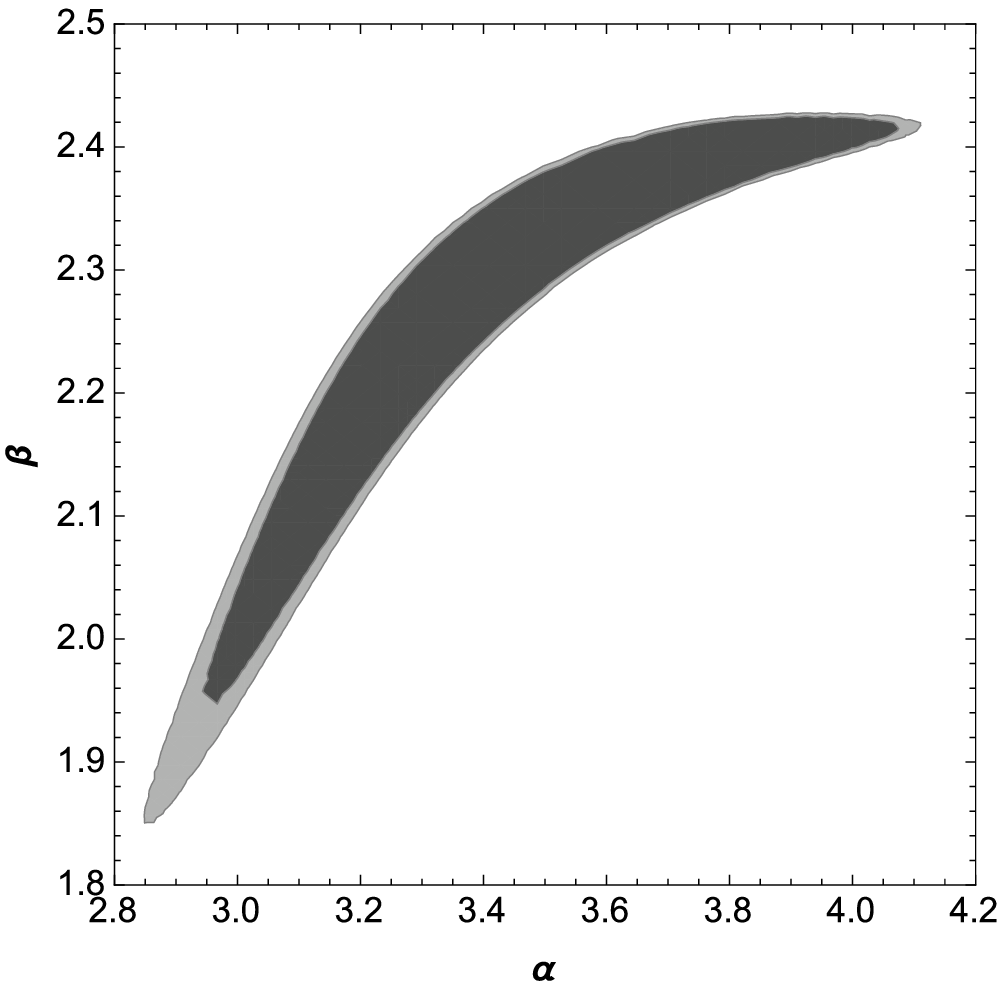}
\caption{ This figure shows the plots for $1\protect\sigma $ (dark shaded)
and $2\protect\sigma $ (light shaded) likelihood contours in the $\protect%
\alpha -\protect\beta $ planes, obtained for model-XI$^{\ast }$ \& XII$%
^{\ast }$.}
\label{fig:contour-XII}
\end{figure}

\newpage

The best fit values of $\alpha $ and $\beta $ obtained by minimizing the
chi-square ($\chi ^{2}$) with $1\sigma $ error are obtained as

\begin{center}
\begin{tabular}{lcc}
& $\alpha $ & $\beta $ \\ 
Models--XI$^{\ast }$ & $3.051_{-0.34}^{+0.45}$ & $2.0_{-0.35}^{+0.31}$ \\ 
Models--XII$^{\ast }$ & $3.006_{-0.075}^{+1.05}$ & $2.0_{-0.045}^{+0.42}$%
\end{tabular}%
.
\end{center}

With these values of $\alpha $ and $\beta $, we plot $q(z)$ vs $z$ (see
figure-4) and $w(z)$ vs $z$ (see figure-5) for models--XI$^{\ast }$ \& XII$%
^{\ast }$ for flat ($k=0$) case. For both the models EoS parameter $w(z)$
crosses the phantom divide line in near future. The transition redshift $%
z_{tr}$ from deceleration to acceleration and the redshift $z_{ph}$ at which 
$w(z)$ crosses the phantom divide line are also shown. From the figures we
can find $z_{tr}\approx 0.73$ for model-XI$^{\ast }$ and $z_{tr}\approx 0.58$
for model-XII$^{\ast }$. Similarly, the redshift at which $w(z)$ crosses the
phantom divide line is $z_{ph}\approx -0.38$ for model-XI$^{\ast }$ and $%
z_{ph}\approx -0.01$ for model-XII$^{\ast }$. 

\begin{figure}[tbph]
\centering
\includegraphics[width=0.45\textwidth]{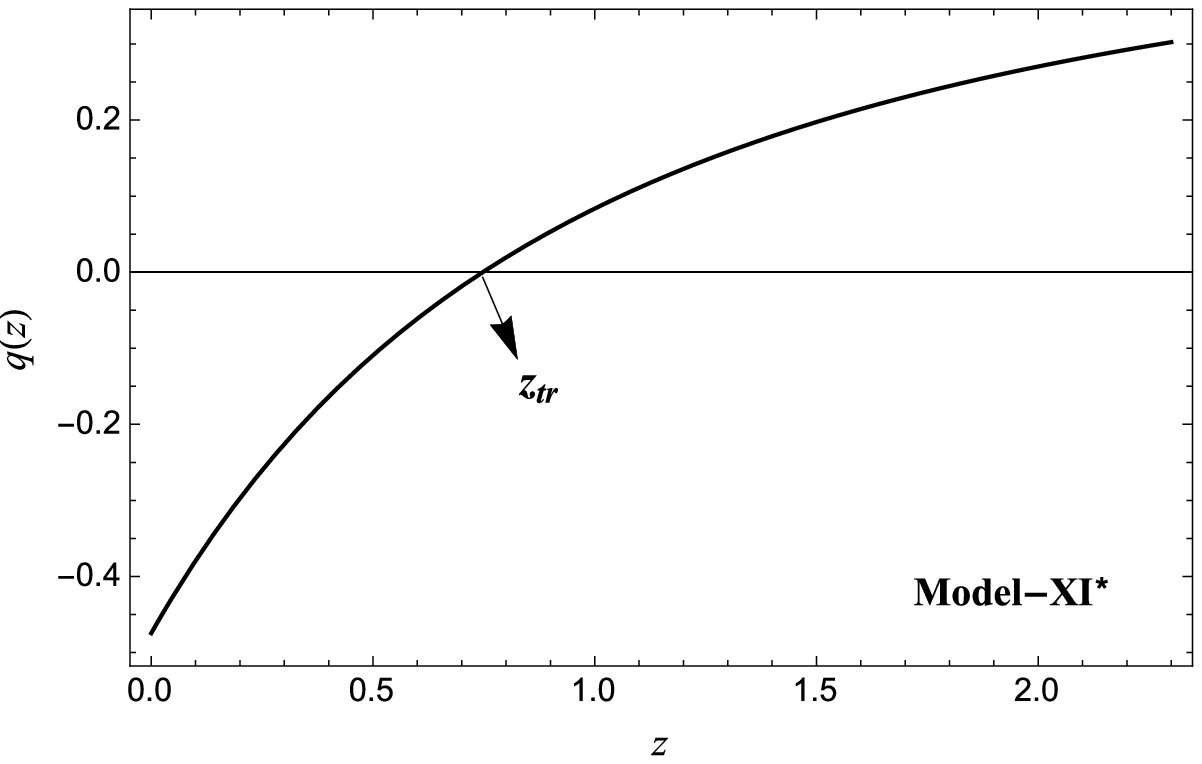} %
\includegraphics[width=0.45\textwidth]{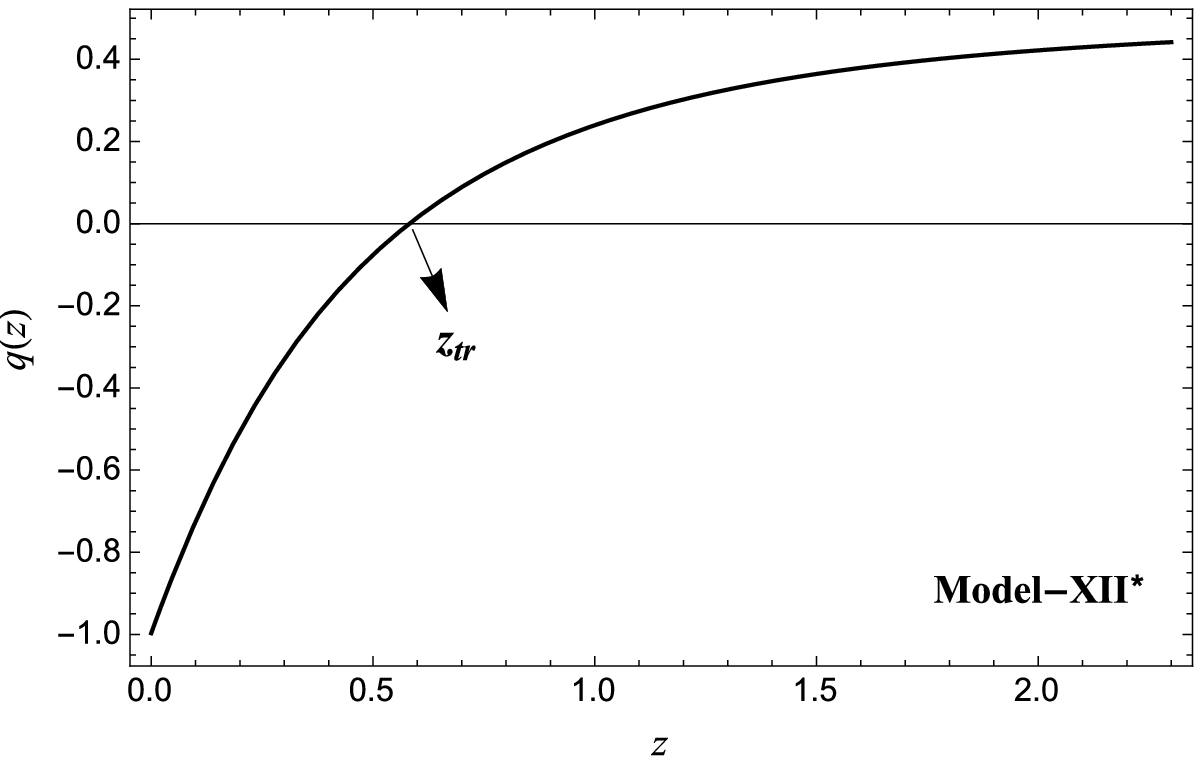}
\caption{This figure shows plots for the DP $q(z)$ vs redshift $z$ for
model-XI$^{\ast }$ \& XII$^{\ast }$ showing the transition from deceleration to acceleration.
}
\label{fig:qz-starXII}
\end{figure}
\begin{figure}[tbph]
\centering
\includegraphics[width=0.45\textwidth]{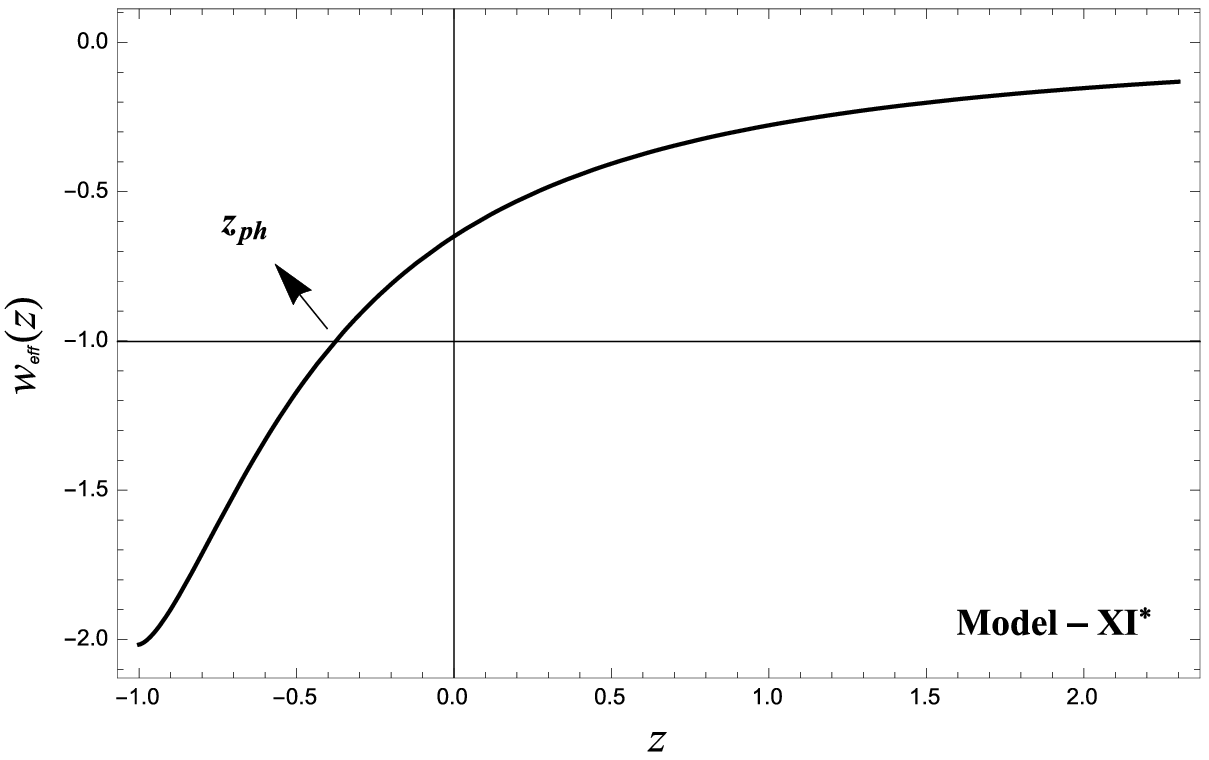} \includegraphics[width=0.45%
\textwidth]{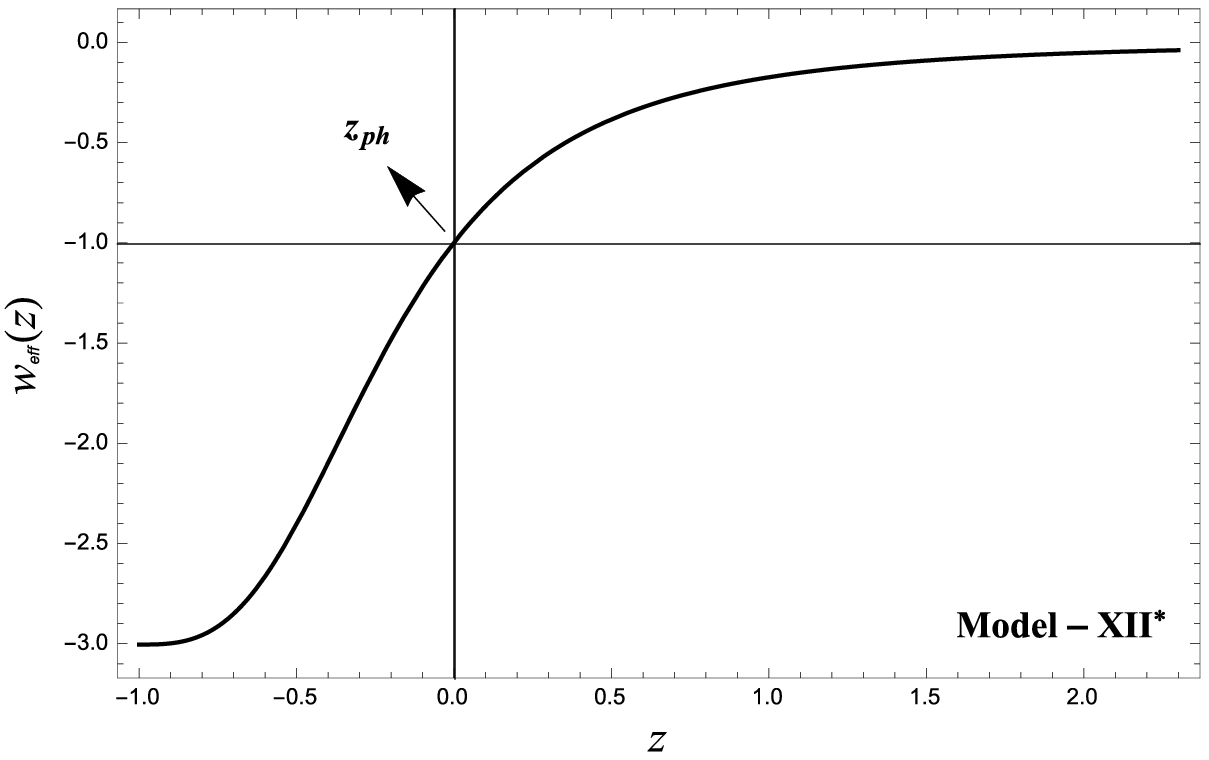}
\caption{This figure shows plots for the EoS $w(z)$ vs redshift $z$ for
model-XI$^{\ast }$ \& XII$^{\ast }$ showing the phantom divide line
crossing. }
\end{figure}

\section{Conclusion}

\qquad In this paper we proposed a convenient and simple parametrization of $%
H$. For certain choices of model parameters in our scheme, we reproduce
several known solutions such as $\Lambda $CDM cosmology, Power law
cosmology, Berman's model, Aksrsu's model, Abdel Rahman's model and others.
Thus, our parametrization covers all these models and also produces some new
cosmologies. The models under consideration either show transition from
deceleration to acceleration or vice versa; in some cases we observed
eternal acceleration. The various constrain equations and parametrization
related to $a(t),$ $H(t),$ $q(t),$ $\Lambda (t),$ $\rho (t),$ $w(t)$
considered in literature are also summarized in detail. As the present
observations agree with deceleration to acceleration transition, we have
analyzed three obtained models which exhibit this important feature. Our
analysis shows that the model-VI has a poor fit for higher redshifts, but
models-XI$^{\ast }$ \& XII$^{\ast }$ show a nice fit with the Hubble data.
The likelihood contours in the $\alpha -\beta $ plane obtained for models--XI%
$^{\ast }$ \& XII$^{\ast }$ are shown in figure-3. The best fit values of $%
\alpha $ and $\beta $ are given by, $\alpha =3.051_{-0.34}^{+0.45}$ \& $%
\beta =2.0_{-0.35}^{+0.31}$ for models-XI$^{\ast }$ and $\alpha
=3.006_{-0.075}^{+1.05}$ \& $\beta =2.0_{-0.045}^{+0.42}$ for models-XII$%
^{\ast }$. For these values of $\alpha $ and $\beta $, the evolution of $q(z)
$ showing the deceleration to acceleration phase transition and the
evolution of $w(z)$ showing the phantom divide line crossing for these
models are shown in figure-4 and figure-5 respectively. 

\qquad It is interesting to note that our parametrization for $H$ can give
rise to several interesting features and can further be studied in
anisotropic Bianchi space-times in the framework of general theory of
relativity as well as in modified theories of gravity. In particular it
would be interesting to discuss the problem of future singularities in the
proposed framework which we deffer to our future investigations.

\bigskip {\Large Acknowledgement} \textit{The authors wish to thank M. Sami
for his useful comments and suggestions throughout the work. Author SKJP
wish to thank Department of Atomic Energy (DAE), Government of India for
financial support through the post-doctoral fellowship of National Board of
Higher Mathematics (NBHM). SKJP is also thankful to J. V. Narlikar, S.
Jhingan and R. G. Vishwakarma for discussions on the related theme and
thankful to Safia Ahmad and Dharm Veer Singh for their help in some
numerical computations.}

{\Large Appendix-1}

\begin{tabular}{lc}
{\small Table-6} &  \\ \hline
\multicolumn{1}{|l}{{\small Variations of Scale factor (}$a${\small )}} & 
\multicolumn{1}{|c|}{\small Ref.} \\ \hline
\multicolumn{1}{|l}{${\small a\sim t}^{n}$} & \multicolumn{1}{|c|}{{\small 
\cite{PLC1}, \cite{PLC2}}} \\ \hline
\multicolumn{1}{|l}{${\small a\sim }\exp {\small (\beta t)}$} & 
\multicolumn{1}{|c|}{{\small \cite{Ellismadsen}}} \\ \hline
\multicolumn{1}{|l}{${\small a\sim }\sinh {\small (\beta t)}$, ${\small %
a\sim }\cosh {\small (\beta t)}$} & \multicolumn{1}{|c|}{{\small \cite%
{Ellismadsen}}} \\ \hline
\multicolumn{1}{|l}{${\small a\sim }\left( \frac{t}{t_{0}}\right) ^{\alpha }%
{\small e}^{\beta \left( \frac{t}{t_{0}}-1\right) }$} & \multicolumn{1}{|c|}{%
{\small \cite{HSF1}}} \\ \hline
\multicolumn{1}{|l}{${\small a\sim e}^{\alpha t}{\small t}^{n}$} & 
\multicolumn{1}{|c|}{{\small \cite{HSF2}}} \\ \hline
\multicolumn{1}{|l}{${\small a(t)=e}^{\alpha (t-t_{s})^{2(1+\beta )}}$} & 
\multicolumn{1}{|c|}{{\small \cite{odintsov1}}} \\ \hline
\multicolumn{1}{|l}{${\small a(t)=a}_{0}\left( \frac{t}{t_{s}-t}\right)
^{\gamma }$} & \multicolumn{1}{|c|}{{\small \cite{odintsov2}}} \\ \hline
\end{tabular}%
\begin{tabular}[t]{l}
{\small where }${\small n}${\small , }${\small \alpha }${\small ,} ${\small %
\beta }${\small ,} ${\small \gamma \ }${\small are constants.} \\ 
${\small t}_{0}${\small \ is the present time.} \\ 
$t_{s}${\small \ future singularity time.}%
\end{tabular}

\begin{tabular}{lc}
{\small Table-7} &  \\ \hline
\multicolumn{1}{|l}{%
\begin{tabular}{l}
{\small Variations of} \\ 
{\small Energy density (}$\rho ${\small )}%
\end{tabular}%
} & \multicolumn{1}{|c|}{\small Ref.} \\ \hline
\multicolumn{1}{|l}{${\small \rho =\rho }_{c}$} & \multicolumn{1}{|c|}{%
{\small \cite{OT1}, \cite{OT2}, \cite{ABDEL}}} \\ \hline
\multicolumn{1}{|l}{${\small \rho \sim \theta }^{2}$} & \multicolumn{1}{|c|}{%
{\small \cite{SKJP1}, \cite{SKJP2}}} \\ \hline
\multicolumn{1}{|l}{${\small \rho =}\frac{A}{a^{4}}\sqrt{{\small a}^{2}%
{\small +b}}$} & \multicolumn{1}{|c|}{{\small \cite{ISLAM}, \cite{SKJP3}}}
\\ \hline
\multicolumn{1}{|l}{$\left( {\small \rho +3p}\right) {\small a}^{3}{\small =A%
}$} & \multicolumn{1}{|c|}{{\small \cite{ASRGVpramana}, \cite{RGVgrg}}} \\ 
\hline
\multicolumn{1}{|l}{${\small \rho +p=\rho }_{c}$} & \multicolumn{1}{|c|}{%
{\small \cite{ASRGVprd}, \cite{ASRGVaustr}}} \\ \hline
\end{tabular}
\begin{tabular}{l}
$A${\small , }$b${\small \ are constants.}%
\end{tabular}

\begin{tabular}{lcc}
{\small Table-8} &  &  \\ \hline
\multicolumn{1}{|l}{{\small Pressure (}${\small p}${\small ) considerations\ 
}} & \multicolumn{1}{|c}{\small EoS} & \multicolumn{1}{|c|}{\small Ref.} \\ 
\hline
\multicolumn{1}{|l}{${\small p=w\rho }$} & \multicolumn{1}{|c}{\small %
perfect fluid} & \multicolumn{1}{|c|}{} \\ \hline
\multicolumn{1}{|l}{${\small p=w\rho -f(H)}$} & \multicolumn{1}{|c}{\small %
viscous fluid} & \multicolumn{1}{|c|}{{\small \cite{viscous1}, \cite%
{viscous2}}} \\ \hline
\multicolumn{1}{|l}{${\small p=}-{\small \rho -\rho }^{\alpha }$} & 
\multicolumn{1}{|c}{\small DE} & \multicolumn{1}{|c|}{{\small \cite{NOJI-EOS}%
}} \\ \hline
\multicolumn{1}{|l}{${\small p=w\rho +k\rho }^{1+\frac{1}{n}}$} & 
\multicolumn{1}{|c}{\small polytropic} & \multicolumn{1}{|c|}{{\small \cite%
{poly0}, \cite{poly1}}} \\ \hline
\multicolumn{1}{|l}{${\small p=-(w+1)}\frac{\rho ^{2}}{\rho _{P}}{\small %
+w\rho +(w+1)\rho }_{\Lambda }$} & \multicolumn{1}{|c}{\small quadratic} & 
\multicolumn{1}{|c|}{{\small \cite{quad}}} \\ \hline
\multicolumn{1}{|l}{${\small p=-}\frac{B}{\rho }$} & \multicolumn{1}{|c}%
{\small Chaplygin Gas} & \multicolumn{1}{|c|}{{\small \cite{CG1}, \cite{CG2}}
} \\ \hline
\multicolumn{1}{|l}{${\small p=-}\frac{B}{\rho ^{\alpha }}$} & 
\multicolumn{1}{|c}{\small generalized CG} & \multicolumn{1}{|c|}{{\small 
\cite{GCG1}, \cite{GCG2}}} \\ \hline
\multicolumn{1}{|l}{${\small p=A\rho -}\frac{B}{\rho ^{\alpha }}$} & 
\multicolumn{1}{|c}{\small modified CG} & \multicolumn{1}{|c|}{{\small \cite%
{MCG1}, \cite{MCG2}}} \\ \hline
\multicolumn{1}{|l}{${\small p=A\rho -}\frac{B(a)}{\rho ^{\alpha }}$} & 
\multicolumn{1}{|c}{\small variable MCG} & \multicolumn{1}{|c|}{{\small \cite%
{VMCG}}} \\ \hline
\multicolumn{1}{|l}{${\small p=A(a)\rho -}\frac{B(a)}{\rho ^{\alpha }}$} & 
\multicolumn{1}{|c}{\small new variable MCG} & \multicolumn{1}{|c|}{{\small 
\cite{NVMCG}}} \\ \hline
\end{tabular}
\begin{tabular}{l}
${\small \rho }_{P}${\small -Plank density} \\ 
${\small \rho }_{\Lambda }${\small -vacuum density} \\ 
${\small 0\leqslant w\leqslant 1}$, ${\small k\leqslant 0}$, \\ 
${\small A>0},${\small \ }${\small B>0}${\small \ are} \\ 
{\small constants.} \\ 
${\small A(a)}${\small \ \& }${\small B(a)}${\small \ are} \\ 
{\small functions of scale} \\ 
{\small factor.}%
\end{tabular}

{\small 
\begin{tabular}{l}
\textbf{Note:} In literature there exist numerous solutions to Einstein
field equations with the ansatz \\ 
$\sigma ^{2}\varpropto \theta ^{2}$, where $\sigma $\ is the energy density
associated with anisotropy and $\theta $\ is the volume expansion \\ 
scalar in homogeneous anisotropic Bianchi models.%
\end{tabular}%
}

\begin{tabular}{lc}
{\small Table-9} &  \\ \hline
\multicolumn{1}{|l}{{\small Variations of cosmological constant (}$\Lambda $)
} & \multicolumn{1}{|c|}{\small Ref.} \\ \hline
\multicolumn{1}{|l}{${\small \Lambda \sim a}^{-n}$} & \multicolumn{1}{|c|}{%
{\small \cite{var-lam1}, \cite{var-lam2}, \cite{RGVcqg1}}} \\ \hline
\multicolumn{1}{|l}{${\small \Lambda \sim H}^{n}$} & \multicolumn{1}{|c|}{%
{\small \cite{COOPER}, \cite{RGVcqg1}, \cite{AIA}}} \\ \hline
\multicolumn{1}{|l}{${\small \Lambda \sim \rho }$} & \multicolumn{1}{|c|}{%
{\small \cite{RGVcqg1}, \cite{SRAY}, \cite{ArbabLAM1}}} \\ \hline
\multicolumn{1}{|l}{${\small \Lambda \sim t}^{n}$} & \multicolumn{1}{|c|}{%
{\small \cite{COOPER}, \cite{RGVcqg1}}} \\ \hline
\multicolumn{1}{|l}{${\small \Lambda \sim q}^{n}$} & \multicolumn{1}{|c|}{%
{\small \cite{COOPER}}} \\ \hline
\multicolumn{1}{|l}{${\small \Lambda }\sim e^{-\beta a}$} & 
\multicolumn{1}{|c|}{{\small \cite{SGRAJ}}} \\ \hline
\multicolumn{1}{|l}{${\small \Lambda =\Lambda (T)}${\small , }${\small T}$%
{\small \ is Temperature}} & \multicolumn{1}{|c|}{{\small \cite{LINDE}}} \\ 
\hline
\multicolumn{1}{|l}{${\small \Lambda }\sim {\small C+}e^{-\beta t}$} & 
\multicolumn{1}{|c|}{{\small \cite{var-lam3}}} \\ \hline
\multicolumn{1}{|l}{${\small \Lambda =3\beta H}^{2}{\small +\alpha a}^{-2}$}
& \multicolumn{1}{|c|}{{\small \cite{CARVALHO}, \cite{Arbab1}}} \\ \hline
\multicolumn{1}{|l}{${\small \Lambda =\beta }\frac{\ddot{a}}{a}$} & 
\multicolumn{1}{|c|}{{\small \cite{RAWAF}, \cite{SRAY}}} \\ \hline
\multicolumn{1}{|l}{${\small \Lambda =3\beta H}^{2}{\small +\alpha }\frac{%
\ddot{a}}{a}$} & \multicolumn{1}{|c|}{{\small \cite{ArbabLAM2}}} \\ \hline
\multicolumn{1}{|l}{$\frac{d\Lambda }{dt}{\small \sim \beta \Lambda -\Lambda 
}^{2}$} & \multicolumn{1}{|c|}{{\small \cite{var-lammoffat}}} \\ \hline
\end{tabular}
\begin{tabular}[t]{l}
{\small where }${\small n},{\small \alpha ,\beta ,C}${\small \ appearing in}
\\ 
{\small the expressions are constants.} \\ 
{\small For a complete set of decay} \\ 
{\small laws of }${\small \Lambda }${\small \ one can see \cite{COOPER}.}%
\end{tabular}

\begin{tabular}{lc}
{\small Table-10} &  \\ \hline
\multicolumn{1}{|l}{{\small Parametrization of DP (}$q${\small )}} & 
\multicolumn{1}{|c|}{\small Ref.} \\ \hline
\multicolumn{1}{|l}{${\small q=m-1}$} & \multicolumn{1}{|c|}{{\small \cite%
{BERMAN1}, \cite{BERMAN2}}} \\ \hline
\multicolumn{1}{|l}{${\small q(t)=-\alpha t+m-1}$} & \multicolumn{1}{|c|}{%
{\small \cite{AKARSU}}} \\ \hline
\multicolumn{1}{|l}{${\small q(t)=-}\frac{\alpha }{t^{2}}{\small +\beta -1}$}
& \multicolumn{1}{|c|}{{\small \cite{q-ASSRP}}} \\ \hline
\multicolumn{1}{|l}{${\small q(a)=-1-}\frac{\alpha a^{\alpha }}{1+a^{\alpha }%
}$} & \multicolumn{1}{|c|}{{\small \cite{q-BANER1}}} \\ \hline
\multicolumn{1}{|l}{${\small q(z)=q}_{0}{\small +q}_{1}{\small z}$} & 
\multicolumn{1}{|c|}{{\small \cite{q-riess}, \cite{q-chunalima}, \cite%
{q-chuna}, \cite{q-nair}}} \\ \hline
\multicolumn{1}{|l}{${\small q(z)=q}_{0}{\small +q}_{1}{\small z(1+z)}^{-1}$}
& \multicolumn{1}{|c|}{{\small \cite{q-nair}, \cite{q-santos}, \cite%
{q-Xu2008}}} \\ \hline
\multicolumn{1}{|l}{${\small q(z)=q}_{0}{\small +q}_{1}{\small z(1+z)(1+z}%
^{2}{\small )}^{-1}$} & \multicolumn{1}{|c|}{{\small \cite{q-sdas}}} \\ 
\hline
\multicolumn{1}{|l}{${\small q(z)=}\frac{1}{2}{\small +q}_{1}{\small (1+z)}%
^{-2}$} & \multicolumn{1}{|c|}{{\small \cite{q-nair}}} \\ \hline
\multicolumn{1}{|l}{${\small q(z)=q}_{0}{\small +q}_{1}{\small [1+}\ln 
{\small (1+z)]}^{-1}$} & \multicolumn{1}{|c|}{{\small \cite{q-Xu2008}}} \\ 
\hline
\multicolumn{1}{|l}{${\small q(z)=}\frac{1}{2}{\small +(q}_{1}{\small z+q}%
_{2}{\small )(1+z)}^{-2}$} & \multicolumn{1}{|c|}{{\small \cite{q-gong2006}, 
\cite{q-gong2007}, \cite{JLu2011}}} \\ \hline
\multicolumn{1}{|l}{${\small q(z)=-1+}\frac{3}{2}\left( \frac{\left(
1+z\right) ^{q_{2}}}{q_{1}+(1+z)^{q_{2}}}\right) $} & \multicolumn{1}{|c|}{%
{\small \cite{q-campo}}} \\ \hline
\multicolumn{1}{|l}{${\small q(z)=-}\frac{1}{4}\left[ 3q_{1}+1-3(q_{1}+1)%
\left( \frac{q_{1}e^{q_{2}\left( 1+z\right) }-e^{-q_{2}\left( 1+z\right) }}{%
q_{1}e^{q_{2}\left( 1+z\right) }+e^{-q_{2}\left( 1+z\right) }}\right) \right]
$} & \multicolumn{1}{|c|}{{\small \cite{q-pavon}}} \\ \hline
\multicolumn{1}{|l}{${\small q(z)=-}\frac{1}{4}{\small +}\frac{3}{4}\left( 
\frac{q_{1}e^{q_{2}\frac{z}{\sqrt{1+z}}}-e^{-q_{2}\frac{z}{\sqrt{1+z}}}}{%
q_{1}e^{q_{2}\frac{z}{\sqrt{1+z}}}+e^{-q_{2}\frac{z}{\sqrt{1+z}}}}\right) $}
& \multicolumn{1}{|c|}{{\small \cite{q-pavon}}} \\ \hline
\multicolumn{1}{|l}{${\small q(z)=q}_{f}{\small +}\frac{q_{i}-q_{f}}{1-\frac{%
q_{i}}{q_{f}}\left( \frac{1+z_{t}}{1+z}\right) ^{\frac{1}{\tau }}}$} & 
\multicolumn{1}{|c|}{{\small \cite{q-ishida}}} \\ \hline
\multicolumn{2}{l}{${\small m}$, ${\small \alpha }$, ${\small \beta }$, $%
{\small q}_{0}$, ${\small q}_{1}$, ${\small q}_{2}${\small \ appearing in
the above expressions are constants.}}%
\end{tabular}

\begin{tabular}{lcl}
{\small Table-11} &  &  \\ \hline
\multicolumn{1}{|l}{{\small Parametrization of EoS (}$w${\small )}} & 
\multicolumn{1}{|c}{\small Ref.} & \multicolumn{1}{|l|}{} \\ \hline
\multicolumn{1}{|l}{${\small w(z)=w}_{0}{\small +w}_{1}{\small z}$} & 
\multicolumn{1}{|c}{{\small \cite{w-LIN1}, \cite{w-LIN2}}} & 
\multicolumn{1}{|l|}{\small Linear} \\ \hline
\multicolumn{1}{|l}{${\small w(z)=w}_{0}{\small +w}_{1}\frac{z}{\left(
1+z\right) ^{2}}$} & \multicolumn{1}{|c}{{\small \cite{w-JBP}}} & 
\multicolumn{1}{|l|}{\small JBP} \\ \hline
\multicolumn{1}{|l}{${\small w(z)=w}_{0}{\small +w}_{1}\frac{z}{1+z}$} & 
\multicolumn{1}{|c}{{\small \cite{w-CPL1}, \cite{w-CPL2}}} & 
\multicolumn{1}{|l|}{\small CPL} \\ \hline
\multicolumn{1}{|l}{${\small w(z)=w}_{0}{\small +w}_{1}\frac{z}{\sqrt{1+z^{2}%
}}$} & \multicolumn{1}{|c}{{\small \cite{w-sqrt}}} & \multicolumn{1}{|l|}%
{\small sqrt} \\ \hline
\multicolumn{1}{|l}{${\small w(z)=w}_{0}{\small +w}_{1}\frac{z(1+z)}{1+z^{2}}
$} & \multicolumn{1}{|c}{{\small \cite{w-BA}}} & \multicolumn{1}{|l|}{\small %
BA} \\ \hline
\multicolumn{1}{|l}{${\small w(z)=w}_{0}{\small +w}_{1}\frac{z}{1+z^{2}}$} & 
\multicolumn{1}{|c}{{\small \cite{w-FSSL}}} & \multicolumn{1}{|l|}{\small %
FSLL Model 1} \\ \hline
\multicolumn{1}{|l}{${\small w(z)=w}_{0}{\small +w}_{1}\frac{z^{2}}{1+z^{2}}$%
} & \multicolumn{1}{|c}{{\small \cite{w-FSSL}}} & \multicolumn{1}{|l|}%
{\small FSLL model 2} \\ \hline
\multicolumn{1}{|l}{${\small w(z)=w}_{0}{\small +w}_{1}\sin {\small (z)}$} & 
\multicolumn{1}{|c}{{\small \cite{w-sin}}} & \multicolumn{1}{|l|}{\small Sine%
} \\ \hline
\multicolumn{1}{|l}{${\small w(z)=w}_{0}{\small +w}_{1}\ln {\small (1+z)}$}
& \multicolumn{1}{|c}{{\small \cite{w-LOG}}} & \multicolumn{1}{|l|}{\small %
Logarithmic} \\ \hline
\multicolumn{1}{|l}{${\small w(z)=w}_{0}{\small +w}_{1}\left( \frac{\ln (2+z)%
}{1+z}-\ln 2\right) $} & \multicolumn{1}{|c}{{\small \cite{w-MZ}}} & 
\multicolumn{1}{|l|}{\small MZ Model 1} \\ \hline
\multicolumn{1}{|l}{${\small w(z)=w}_{0}{\small +w}_{1}\left( \frac{\sin
(1+z)}{1+z}-\sin 1\right) $} & \multicolumn{1}{|c}{{\small \cite{w-MZ}}} & 
\multicolumn{1}{|l|}{\small MZ Model 2} \\ \hline
\multicolumn{1}{|l}{${\small w(z)=w}_{0}{\small +w}_{1}\left( \frac{z}{1+z}%
\right) ^{n}$} & \multicolumn{1}{|c}{{\small \cite{w-nCPL}}} & 
\multicolumn{1}{|l|}{\small nCPL} \\ \hline
\multicolumn{1}{|l}{${\small w(z)=w}_{0}{\small +w}_{1}\frac{z}{\left(
1+z\right) ^{n}}$} & \multicolumn{1}{|c}{{\small \cite{w-nCPL}}} & 
\multicolumn{1}{|l|}{\small nJBP} \\ \hline
\multicolumn{1}{|l}{${\small w(z)=w}_{0}{\small +w}_{1}\ln \left( {\small 1+%
\frac{z}{1+z}}\right) $} & \multicolumn{1}{|c}{{\small \cite{w-feng}}} & 
\multicolumn{1}{|l|}{\small Modified Logarithmic} \\ \hline
\multicolumn{3}{l}{${\small w}_{0}$, ${\small w}_{1}${\small \ appearing in
the above expressions are constants.}}%
\end{tabular}


\begin{thebibliography}{999}
\bibitem{sandage} A. Sandage, Physics Today 34 (1970).

\bibitem{SCP} S. Perlmutter et al., Astrophy. J. 517 565 (1999).

\bibitem{HZTEAM} A. G. Reiss et al., Astron. J. 116, 1009 (1998).

\bibitem{HZ1} A. G. Riess et al., Astrophy. J. 536 62 (2000).

\bibitem{HZ2} A. G. Riess et al., Astrophy. J. 659 98 (2007).

\bibitem{HZ3} D. H. Weinberg et al., Phys. Rept. 530 87 (2013).

\bibitem{SCP1} P. Astier et al., Astron. \& Astrophy. 447 31 (2006).

\bibitem{SCP2} Amanullah, et al., Astrophy. J. (2010).

\bibitem{SCP3} D. Rubin et al., Astrophy. J. (2013).

\bibitem{cmb1} A. H. Jaffe et al., Phys. Rev. Lett. 86 3475 (2001).

\bibitem{cmb2} D. N. Spergel et al., Astrophys. J. Suppl. 170 377 (2007).

\bibitem{bao1} J. R. Bond et al. Mon. Not. R. Astron. Soc. 291 L33 (1997).

\bibitem{bao2} Y. Wang and P. Mukherjee, Astrophy. J. 650 1 (2006).

\bibitem{sdss1} U. Seljak et al., Phys. Rev. D 69 103501 (2004).

\bibitem{sdss2} J. K. Adelman-McCarthy et al., Astrophy. J. Suppl. 162 38
(2005).

\bibitem{vishwa0} R. G. Vishwakarma, arXiv: 1605.09236v1 (2016).

\bibitem{vishwa1} R. G. Vishwakarma, Int. J. Geom. Meth. Mod. Phys. 12
1550116 (2015).

\bibitem{vishwa2} R. G. Vishwakarma, Frontiers of Physics 9(1) 98 (2014).

\bibitem{vishwa3} R. G. Vishwakarma, Res. Astron. Astrophys. 13 1409 (2013).

\bibitem{sahniCC} V. Sahni, Int. J. Mod. Phys. D 9 373 (2000).

\bibitem{vishwaCC} R. G. Vishwakarma, Mon. Not. R. Astron. Soc. 331 776
(2002).

\bibitem{peeblesCC} P. J. E. Peebles and B. Ratra, Rev. Mod. Phys. 75 559
(2003).

\bibitem{WEINBERG} S. Weinberg, Rev. Mod. Phys. 61 1 (1989).

\bibitem{quint1} B. Ratra and P. J. E. Peebles, Phys. Rev. D 37, 3406 (1988).

\bibitem{quint2} R. R. Caldwell, R. Dave and P. J. Steinhardt, Phys. Rev.
Lett. 80 1582 (1998).

\bibitem{quint3} V. Sahni, M. Sami and T. Souradeep, Phys. Rev. D 65 023518
(2002).

\bibitem{quint4} M. Sami and T. Padmanabhan, Phys. Rev. D 67 083509 (2003).

\bibitem{kessen1} Armendariz-Picon, T. Damour and V. Mukhanov, Phys. Lett. B
458 209 (1999).

\bibitem{kessen2} T. Chiba, T. Okabe and M. Yamaguchi, Phys. Rev. D 62
023511 (2000).

\bibitem{spint} L. A. Boyle, R. R. Caldwell and M. Kamionkowski, Phys. Lett.
B 545 17 (2002).

\bibitem{tachy1} A. Sen, J. High Energy Phys. 0207 065 (2002).

\bibitem{tachy2} T. Padmanabhan, Phys. Rev. D 66 021301 (2002).

\bibitem{quintom1} B. Feng, X. L. Wang and X. M. Zhang, Phys. Lett. B 607 35
(2005).

\bibitem{quintom2} Z. K. Guo, Y. S. Pio, Y. Z. Zhang and X. M. Zhang, Phys.
Lett. B 608 177 (2005).

\bibitem{quintom3} M. R. Setare, J. Sadeghi and A. R. Amani, Phys. Lett. B
660 299 (2008).

\bibitem{quintom4} M. R. Setare, E. N. Saridakis, J. Cosmol. Astropart.
Phys. 09 026 (2008).

\bibitem{chamel1} J. Khoury, and A. Weltman, Phys. Rev. Lett. 93 171104
(2004).

\bibitem{chamel2} P. Brax, et al., Phys. Rev. D 70 123518 (2004).

\bibitem{chamel3} Abdussattar and S.R. Prajapati, Int. J. Theor. Phys. 50
2355 (2011).

\bibitem{phant1} L. Parker and A. Raval, Phys. Rev. D 60 063512 (1999).

\bibitem{phant2} V. Sahni and A. Starobinsky, Int. J. Mod. Phys. D 9 373
(2000).

\bibitem{phant3} S. Nojiri, S. D. Odintsov, Phys. Lett. B 562 147 (2003).

\bibitem{phantsami} Parampreet Singh, M. Sami, Naresh Dadhich, Phys. Rev. D
68 023522(2003).

\bibitem{phant4} M. Sami and A. Toporensky, Mod. Phys. Lett. A 19 1509
(2004).

\bibitem{phant5} A. V. Astashenok et al., Phys. Lett. B 709(4) 396 (2012).

\bibitem{CG1} A. Yu. Kamenshchik, U. Moschella and V. Pasquier, Phys. Lett.
B 511 265 (2001).

\bibitem{CG2} V. Gorini et al., AIP Conf. Proc. 751 108 (2005).

\bibitem{DEREV1} E. J. Copeland, M. Sami and S. Tsujikawa, Int. J. Mod.
Phys. D 15 1753 (2006).

\bibitem{DEREVa} M. Sami, New Adv. Phys. 10 77 (2016).

\bibitem{DEREVb} M. Sami, R. Myrzakulov, arXiv:1309.4188v2 (2013).

\bibitem{DEREVc} Md. Wali Hossain et al., Int. J. Mod. Phys. D 24 1530014
(2015).

\bibitem{DEREVd} M. Sami, Curr. Sci. 97 887 (2009).

\bibitem{DEREVe} M Sami, arXiv:0901.0756v1 (2009).

\bibitem{DEREV2} J. Yoo and Y. Watanabe, Int. J. Mod. Phys. D 21 1230002
(2012).

\bibitem{Turner} M. S. Turner and A. G. Riess, Astrophys. J. 569 18 (2002).

\bibitem{Weinberg} S. Weinberg, \textit{Cosmology and Gravitation} (John
Wiley Sons, New York) (1972).

\bibitem{LINDER-w} Eric V. Linder, Phys. Rev. D 73 063010 (2006).

\bibitem{BOLOTIN-q} Yu. L. Bolotin et al., arXiv:1502.00811 (2015).

\bibitem{BOLO} Yu. L. Bolotin et al., arXiv:1506.08918 (2015).

\bibitem{H-odin3} S. Nojiri, S. D. Odintsov and V. K. Oikonomou, Phys. Lett.
B 747 310 (2015).

\bibitem{H-BermanNuovo} M. S. Berman, Nuovo Cimento 74(2) (1983).

\bibitem{H-Banerjee1} N Banerjee, S. Das and K. Ganguly, Pramana 74(3) 481
(2010).

\bibitem{H-JPSINGH} J. P. Singh, Astrophys. Space Sci. 318 103 (2008).

\bibitem{H-Pacif1} S. K. J. Pacif and B. Mishra, Res. Astron. Astrophys.
15(12) 2141 (2015).

\bibitem{H-Pacif2} S. K. J. Pacif and B. Mishra, Astrophys. Space Sci. 360
48 (2015).

\bibitem{H-Cannata} F. Cannata and A. Y. Kamenshchik, Int. J. Mod. Phys. D
20 121 (2010).

\bibitem{H-nojiri} S. Noniri and S. D. Odintsov, Gen. Relativ. Grav. 38(8)
1285 (2006).

\bibitem{H-odinF} Paul H. Frampton et al., Phys. Lett. B 708 204 (2012).

\bibitem{H-odin1} S. Nojiri et al., J. Cosmol. Astropart. Phys. 1509 044
(2015).

\bibitem{H-odin2} K. Bamba et al., Phys. Lett. B 732 349 (2014).

\bibitem{PLC1} D. Lohiya and M. Sethi, Class. Quantum Grav. 16 1545 (1999).

\bibitem{BERMAN1} M. S. Berman and F. M. Gomide, Gen Relativ. Grav. 20 191
(1998).

\bibitem{ABDELMODEL} A-M. M. Abdel Rahman, Phys. Rev. D 45(10) 3497 (1992).

\bibitem{q-ASSRP} Abdussattar and S. R. Prajapati, Astrophys. Space Sci. 331
657 (2011).

\bibitem{AKARSU} O. Akarsu and T. Dereli, Int. J. Theoret. Phys. 51 612
(2012).

\bibitem{HSF1} Ozgur Akarsu et al., J. Cosmol. Astropart. Phys. 01 22 (2014).

\bibitem{HSF2} B. Mishra and S. K. Tripathy, Modern Physics Letters A 30(36)
1550175 (2015).

\bibitem{COOPER} J. M. Overduin and F. I. Cooperstock, Phys. Rev. D 58
043506 (1998).

\bibitem{vishwa-metalic} R. G. Vishwakarma, Mon. Not. R. Astron. Soc. 345
545 (2003).

\bibitem{Simon} J. Simon et al., Phys. Rev. D 71 123001 (2005).

\bibitem{Moresco} M. Moresco et al., J. Cosmol. Astropart. Phys. 1208 006
(2012).

\bibitem{Cuesta} J. A. Cuesta et al., Mon. Not. R. Astron. Soc. 457 1770
(2016).

\bibitem{Blake} C. Blake et al., Mon. Not. R. Astron. Soc. 425 405 (2012).

\bibitem{Stern} D. Stern et al., J. Cosmol. Astropart. Phys. 1002 008 (2010).

\bibitem{Moresco2} M. Moresco et al., J. Cosmol. Astropart. Phys. 1605 014
(2016).

\bibitem{Delubac} T. Delubac et al., Astron. \& Astrophys. 574 A59 (2015).

\bibitem{Font} A. Font-Ribera et al., J. Cosmol. Astropart. Phys. 1405 027
(2014).

\bibitem{PLC2} A. Batra, M. Sethi and D. Lohiya, Phys. Rev. D 60 108301
(1999).

\bibitem{Ellismadsen} G. F. R. Ellis and M. S. Madsen, Class. Quantum Grav.
8 667 (1991).

\bibitem{odintsov1} S. D. Odintsov and V. K. Oikonomou, Phys.Rev. D 92(2)
024016 (2015).

\bibitem{odintsov2} S. Nojiri, S. D. Odintsov and S. Tsujikawa, Phys. Rev. D
71 063004 (2005).

\bibitem{OT1} M. Ozer and M.O. Taha, Phys. Lett. B 171 363 (1986).

\bibitem{OT2} M. Ozer and M.O. Taha, Nucl. Phys. B 287 776 (1987).

\bibitem{ABDEL} A-M. M. Abdel-Rahman, Gen. Relativ. Gravit. 22 655 (1990).

\bibitem{SKJP1} S. K. J. Pacif and Abdussattar, Eur. Phys. J. Plus 129 244
(2014).

\bibitem{SKJP2} S. K. J. Pacif and Abdussattar, Proceedings of 3rd
International Conference on \textquotedblleft Innovative Approach in Applied
Physical, Mathematical/Statistical, Chemical Sciences and Emerging Energy
Technology for Sustainable Development\textquotedblright , New Delhi (2014).

\bibitem{ISLAM} J. N. Islam, \textit{An Introduction to Mathematical
Cosmology} (Cambridge University Press, UK) (1992).

\bibitem{SKJP3} Abdussattar and S. R. Prajapati, Chin. Phys. Lett. Vol.
28(2) 029803 (2011).

\bibitem{ASRGVpramana} Abdussattar and R. G. Vishwakarma, Pramana 47 41
(1996).

\bibitem{RGVgrg} R. G. Vishwakarma, Gen. Relativ. Grav. 33(11) 1973 (2001).

\bibitem{ASRGVprd} R. G. Vishwakarma, Abdussattar and A. Beesham, Phys. Rev.
D 60 063507 (1999).

\bibitem{ASRGVaustr} Abdussattar and R. G. Vishwakarma, Aust. J. Phys. 50
893 (1997).

\bibitem{var-lam1} T. S. Olson and T. F. Jordan, Phys. Rev. D 35 3258 (1987).

\bibitem{var-lam2} D. Pavon, Phys. Rev. D 43 375 (1991).

\bibitem{RGVcqg1} R. G. Vishwakarma, Class. Quantum Grav. 18, 1159 (2001).

\bibitem{SRAY} S. Ray et al., Int. J. Theor. Phys. 48(9) 2499 (2009).

\bibitem{AIA} Arbab I Arbab, Gen. Relativ. Grav. 29 61 (1997).

\bibitem{ArbabLAM1} Arbab I Arbab, Chin. Phys. Lett. 25 4497 (2008).

\bibitem{SGRAJ} S. G. Rajeev, Phys. Lett. B 125 144 (1983).

\bibitem{LINDE} A. D. Linde, JETP Lett. 19, 183 (1974).

\bibitem{var-lam3} A. Beesham, Phys. Rev. D 48 3539 (1993).

\bibitem{CARVALHO} J. C. Carvalho, J A S Lima and I Waga, Phys. Rev. D 46
2404 (1992).

\bibitem{Arbab1} Arbab I Arbab and A.-M. M. Abdel-Rahman, Phys. Rev. D
50(12) 7725 (1995).

\bibitem{RAWAF} A. S. Al-Rawaf, Mod. Phys. Lett. A 13 429 (1998).

\bibitem{ArbabLAM2} Arbab I Arbab, Grav. Cosmol. 8 227 (2002).

\bibitem{var-lammoffat} J. W. Moffat, Los Alamos report astro-ph/9608202
(1996).

\bibitem{viscous1} C. Ekart, Phys. Rev. 58 919 (1940).

\bibitem{viscous2} O. Gron, Astrophys. Space Sci. 173 91 (1990).

\bibitem{NOJI-EOS} S. Nojiri and S. D. Odintsov, Phys. Rev. D 70, 103522
(2004).

\bibitem{poly0} K. Karami, S. Ghaari, and J. Fehri, Eur. Phys. J. C 64(1) 85
(2009).

\bibitem{poly1} Pierre-Henri Chavanis, Eur. Phys. J. Plus 129 38 (2014).

\bibitem{quad} Pierre-Henri Chavanis, arXiv:1309.5784v2 (2015).

\bibitem{GCG1} M. C. Bento, O. Bertolami and A. A. Sen, Phys. Rev. D 66
043507 (2002).

\bibitem{GCG2} V. Gorini, A. Yu. Kamenshchik and U. Moschella, Phys. Rev. D
67 063509 (2003).

\bibitem{MCG1} H. B. Benaoum, arxiv:hep-th/0205140v1 (2002)

\bibitem{MCG2} U. Debnath, A. Banerjee, and S. Chakraborty, Class. Quantum
Grav. 21 5609 (2004).

\bibitem{VMCG} U. Debnath, Astrophys. Space Sci. 312 295 (2007).

\bibitem{NVMCG} W. Chakraborty, U. Debnath, Gravit. and Cosmol. 16 223
(2010).

\bibitem{BERMAN2} M. S. Berman, Phys. Rev. D 43 1075 (1991).

\bibitem{q-BANER1} N. Banerjee and S. Das, Gen. Relativ. Grav. 37 1695
(2005).

\bibitem{q-riess} A. G. Riess et al., Astrophy. J. 607 665 (2004).

\bibitem{q-chunalima} J. V. Cunha and J. A. S. Lima, Mon. Not. R. Astron.
Soc. 390 210 (2008).

\bibitem{q-chuna} J. V. Cunha, Phys. Rev. D 79 047301 (2009).

\bibitem{q-nair} R. Nair et al., J. Cosmol. Astropart. Phys. 01 018 (2012).

\bibitem{q-santos} B. Santos, J. C. Carvalho, J. S. Alcaniz, Astropart.
Phys. 35 17 (2011).

\bibitem{q-Xu2008} L. Xu and H. Liu, Mod. Phys. Lett. A 23 1939 (2008).

\bibitem{q-sdas} Abdulla Al Mamon and Sudipta Das, arXiv:1507.00531v1 (2015).

\bibitem{q-gong2006} Y. G. Gong and A. Wang, Phys. Rev. D 73 083506 (2006).

\bibitem{JLu2011} J. Lu, L. Xu and M. Liu, Phys. Lett. B 699 246 (2011).

\bibitem{q-gong2007} Y. G. Gong and A. Wang, Phys. Rev. D 75 043520 (2007).

\bibitem{q-campo} S. del Campo et al., Phys. Rev. D 86 083509 (2012).

\bibitem{q-pavon} D. Pavon et al., Proc. MG13 Meeting Gen. Relativ.,
Stockholm Univ., Sweden (2012). arXiv:1212.6874v1.

\bibitem{q-ishida} E. E. O. Ishida et al., Astroparticle Phys. 28 7 (2007).

\bibitem{w-LIN1} D. Huterer, M. S. Turner, Phys. Rev. D 64 123527 (2001).

\bibitem{w-LIN2} J. Weller and A. Albrecht, Phys. Rev. D 65 103512 (2002).

\bibitem{w-JBP} H. K. Jassal, J. S. Bagla, and T. Padmanabhan, Mon. Not. R.
Astron. Soc. Letters 356(1) L11 (2005).

\bibitem{w-CPL1} M. Chevallier and D. Polarski, Int. J. Mod. Phys. D 10 213
(2001).

\bibitem{w-CPL2} E. V. Linder, Phys. Rev. Lett. 90 091301 (2003).

\bibitem{w-sqrt} G. Pantazis, S. Nesseris and L. Perivolaropoulos, Phys.
Rev. D 93 103503 (2016).

\bibitem{w-BA} E. M. Barboza, Jr. and J. S. Alcaniz, J. Cosmol. Astropart.
Phys. 02 042 (2012).

\bibitem{w-FSSL} C. -J. Feng et al., J. Cosmol. Astropart. Phys. 09 023
(2012).

\bibitem{w-sin} R. Lazkoz, V. Salzano, and I. Sendra, Phys. Lett. B 694 198
(2010).

\bibitem{w-LOG} G. Efstathiou, Mon. Not. R. Astron. Soc. 310 842 (1999).

\bibitem{w-MZ} J. -Z. Ma and X. Zhang, Phys. Lett. B 699 233 (2011).

\bibitem{w-nCPL} Dao-Jun Liu et al., Mon. Not. R. Astron. Soc. 388 275
(2008).

\bibitem{w-feng} L. Feng and T. Lu, J. Cosmol. Astropart. Phys. 11 034
(2011).
\end{thebibliography}
\end{document}